\documentclass[a4paper, 12pt]{article}

\usepackage[utf8]{inputenc}
\usepackage{mathptmx}
\usepackage{mathtools}
\usepackage{latexsym,bm}
\usepackage{amsfonts}
\usepackage{amssymb}
\usepackage{amsmath}
\usepackage{amstext}
\usepackage{graphicx}
\usepackage{adjustbox}
\usepackage{multirow}
\usepackage{multicol}

\begin{document} 
\title{An improved  mathematical model for the sedimentation of microplastic particles in a lid-driven cavity with obstacle}
\author{\small{Nityananda Roy $^{1}$,\quad Thomas Goetz $^2$\thanks{\footnotesize{Corresponding author: goetz@uni-koblenz.de}}, \quad Karunia Putra Wijaya $^2$, \quad S. Sundar $^1$} 
\\[2mm]
\footnotesize{$^{1}$ Department of Mathematics, Indian Institute of Technology Madras, Chennai 600036, India } \\
\footnotesize{$^2$ Mathematical Institute, University of Koblenz, 56070 Koblenz, Germany. }  } 
\date{}
\maketitle

\begin{abstract}
\noindent In this paper, we developed a mathematical model for the sedimentation process of small particles in a lid--driven cavity flow with an obstacle to model the transport of microplastic particles in rivers. A stationary incompressible Navier--Stokes simulation at moderate Reynolds--numbers provides the background flow field. Spherical particles are injected into this flow field and their equation of motion is solved to determine the number of particles that sediment on the surface of the obstacle. To capture the effect of typical biological organisms and biofilms on the bottom surface of river beds, the obstacle can exert an attraction force onto the particles. Various simulations and parameter studies are carried out to determine the influence of the obstacle geometry, particle densities, and the attraction force on the sedimentation process.
\\[2mm]
\textbf{Key words:~~~}Lid--driven cavity; Navier--Stokes equation; micro particles; sedimentation;
\end{abstract}

\section{Introduction}
Microplastic refers to a piece of plastic with the size measured by the longest diameter varying between $1~\mu\text{m}$ and $5~\text{mm}$. The pollution of water reservoirs with microplastic particles has received considerable attention during the last years. According to  \cite{smith2018}, plastic pollution in world oceans is estimated to have reached 270.000 tonnes, or 5.25 trillion pieces. Microplastic particles come from different sources, larger items of plastic debris fragment, and in addition, micro-- and nanoparticles are nowadays added to vast range of every--day products like beauty cremes or toothpaste. Those tiny particles pass through sewer and filtration systems and enter via the rivers, finally into the ocean. In particular, those microparticles consume by the biological organisms and biofilms and, lastly, assembled in the human body (\cite{cole2013microplastic, smith2018}). The sedimentation of  microplastic particles onto living biological organisms and biofilms on river beds is currently under discussion.

\par An adequate scientific assessment of microplastic pollution requires, among others, an understanding of the transport and sedimentation process of those particles in flows. A quite good number of articles and other scientific publications, including some very sophisticated ones, are available on experiments, models and simulations of the transport of particles in flows. Just to name a few recent ones, numerical and experimental results are shown in \cite{venturini2012, afrouzi2015lattice, tsorng2006three, safdari2015lattice, sidik2012use, portela2003, garoosi2015, liu2011numerical, pan2018}. \cite{tsuji1993discrete} described the transport of particles in a horizontal channel flow. \cite{frank1993numerical} studied particles' transport in a horizontal turbulence channel flow. \cite{kosinski2009simulation} simulated the solid particles' motion in a lid-driven cavity by considering two-phase flow. Later on,  \cite{kosinski2010extension} presented not only the particle collision but also accounted the cohesion between particles and agglomerations of particles. \cite{safdari2014lattice} discussed the transport of particles in a three-dimensional lid-driven cavity.

\par In this paper, a mathematical model has been developed by introducing a physical force between particle and obstacle surface in order to observe the effect of the considered force on the sedimentation of the particle. The motivation to introduce the attachment force is to incorporate the physical phenomena observed in forming the biofilm, which forms at the bottom of the river or ocean as an external layer on the surface of the rocks with a certain gradient of density. The closer look in the biofilm is merely behaved like corals with tentacles; it thus might have the capability of catching objects in the vicinity of the accumulation. We assume that the obstacles can attract the particles which are within a distance less than $ \delta $ from its surface, where the $ \delta$ denotes the "region of influence" around the surface of the obstacle. The particle experiences the attachment force only if it is inside the region of influence. The biofilm is distributed uniformly on the surface of the obstacle and a particle can interact with any biofilm as long as it is in the closest proximity to the particle. To avoid particle infiltration, we shall also include a saturation parameter $\epsilon$ accounting for force boundedness.

From the aforementioned, it is clear that the sedimentation of the particles by placing the obstacles in different angles in the lid-driven cavity along with the attachment force model will be very interesting to study. The particles are considered to be very small; therefore, one-way fluid-particle coupling is discussed  here. The outline of this paper is as follows: the mathematical model is discussed in Section 2, the numerical methods are discussed in Section 3, and the numerical results and the conclusions are summarized in the subsequent sections.

\section{Mathematical modeling}
\begin{figure}[h!]
  \centering
  \includegraphics[width=.5\linewidth]{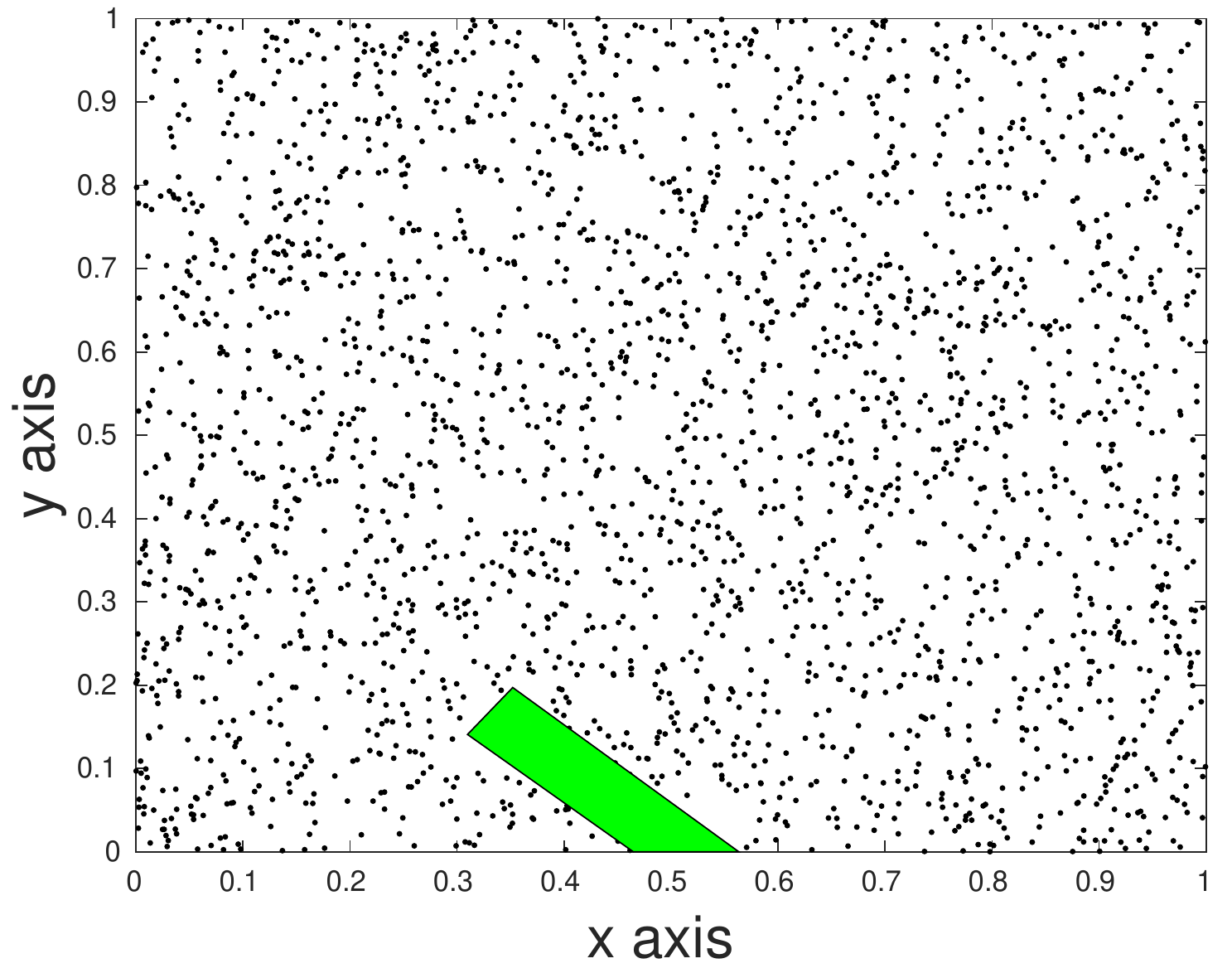}
  \caption{Two--dimensional cross--sectional plane of the cavity with an obstacle (green) and randomly distributed particles.}
  \label{fig:1}
\end{figure}

We consider the sedimentation process of small particles immersed in fluid. The flow domain is given by a long cavity with square cross--sectional area. In order to reduce the computational complexity, we just model the flow in the two--dimensional cross--sectional plane driven by a constant velocity at the top--boundary of the cavity (lid--driven cavity). Moreover, the flow field is considered to be stationary, incompressible and isothermal. Inside the cavity an obstacle disturbs the primary vortex of the flow--field and the particles can sediment on this obstacle, see Figure~\ref{fig:1}. The particles themselves are considered to be small and light; hence we only consider a one way--coupling between flow and particles. The back--coupling of the particles onto the flow field is neglected.  

Let $U=(U_x, U_y)$ denote the flow velocity and $P$ denote the pressure. Incompressibility leads to
\begin{subequations}
\begin{align}
	\nabla\cdot U &= 0
\intertext{and momentum conservation leads to the classical stationary Navier--Stokes equation}
	(U\cdot \nabla)\, U &=-\frac{1}{\rho_f}\nabla P+\nu \Delta U\;.
\label{eq:NS}
\end{align}
\end{subequations}
Here, $\rho_f$ and $\nu$ denote the density and kinematic viscosity of the fluid. In Table~\ref{tab:1} we give values for these and other parameters resembling water at an ambient temperature of $20\,^\circ\,\mathrm{C}$. 

The above PDE--system is supplemented by suitable boundary conditions. To model the lid--driven cavity, we assume at the top boundary a constant horizontal velocity $U=(U_\infty, 0)$ and at the other boundaries we impose a no--slip condition, i.e.,~$U=(0,0)$. 

Each of the particles $\mathcal{X}_p$, $p=1,\dots n$ with position $X_p$, velocity $V_p$ and angular velocity $W_p$ has translational and rotational degrees of freedom. Its equation of motion is given by
\begin{subequations}
\begin{align}
	\frac{d X_p}{dT} &= V_p\;,\\
	m_p \frac{d V_p}{dT} &= -m_p g+\frac{\rho_f}{\rho_p} m_p g 
		+ \frac{1}{2}\rho_f c_d A_p \cdot \left| U-V_p \right| \cdot (U-V_p)+F_a(X_p)\;.\label{eq4} \\
\intertext{The first two terms on the right-hand side of equation \eqref{eq4} account for gravity and buoyancy. The third term represents the drag force depending on the relative velocity between particle and fluid. The fourth term $F_{a}$ models an attachment force between particle and obstacle; details are given below in equation~\eqref{E:Fattach}. The rotational motion of the particle is given by}
	I_p \frac{d W_p}{dT} &= \frac{\rho_f}{2}r_p^5\, c_R \cdot \left| \frac{1}{2}\nabla\times U -W_p\right| \cdot \left(\frac{1}{2}\nabla\times U -W_p\right)\;.
\label{eq5}
\end{align}
\end{subequations}
The mass, density, radius, cross sectional area and moment of inertia of the $p$--th particle are given by $m_p$, $\rho_p$, $r_p$, $A_p$ and $I_p$. The coefficients $c_d$ and $c_R$ denote the translational and rotational drag coefficients. Following~(\cite{schwarzkopf2011multiphase, kosinski2009simulation}), we model these drag--coefficients depending on the according Reynolds number
\begin{subequations}
\begin{align}
	c_d &= \frac{24}{Re_{p}}\;, \\
	c_R &= \begin{cases}
		\frac{64}{Re_r}, & Re_r \le 32\;,\\
		\frac{12.4}{Re_r^{1/2}}+\frac{128.4}{Re_r}, & 32< Re_r \le 1000\;,
	\end{cases}
\end{align}
\end{subequations}
where $Re_{p}= \frac{2 r_{p}\left| U-V_p \right|}{\nu}$ denotes the particle Reynolds number and
\begin{equation*}
	Re_r =\frac{(2r_p)^2 \left|\frac{1}{2}\nabla\times U -W_p\right|}{\nu} 
\end{equation*}
is called the local rotational Reynolds number.

The attachment force $F_a$ models an attraction between the obstacle $\Omega$ and particles in its vicinity (distance less than a threshold value $\delta$). If the force is directed normal to the surface of the obstacles, we may use the relation
\begin{equation}
\label{E:Fattach}
	F_{a} = \begin{cases}
		-K \frac{r_p^2}{{\Vert d\Vert}^2+\epsilon^{2}}\frac{d}{\Vert d\Vert}\,, &\Vert d\Vert \le \delta, \\
		0,	& else,
		\end{cases}
\end{equation}
where $K$ denotes some constant and $d=x_p-y$ is the distance vector between the particles center of mass $x_p$ and the nearest point $y$ of the obstacle, i.e.,~$y = \text{argmin}_{z\in \partial \Omega} \Vert x_p-z\Vert$.

\par In order to simplify the equations we non--dimensionalize them using the characteristic length and velocity $L_\infty$ and $U_\infty$. The characteristic time is given by $T_\infty = L_\infty/U_\infty$ and the pressure differences are scaled using the stagnation pressure $P_0-P_\infty = \rho_f U_\infty^2$. This data gives the usual Reynolds number $Re=U_\infty L_\infty / \nu$. With these scalings, we introduce the dimensionless quantities $u^{+}=U/U_\infty$, $x^{+}=X/L_\infty$, $t^{+}=T/T_\infty$, $v_p^{+}=V_p/U_\infty$, $r_p^{+}=r_p/L_\infty$ and $\omega_p^{+}=T_\infty W_p$. The non--dimensionalized equations read as
\begin{subequations}
\begin{align}
	\nabla^{+}\cdot u^{+} &= 0, \label{E:NS1}\\
	(u^{+}\cdot \nabla^{+})\, u^{+} &=-\nabla^{+} p^{+}+\frac{1}{Re} \Delta^{+} u^{+}, \label{E:NS2}\\
	\frac{d x_p^{+}}{dt^{+}} &= v_p^{+}\;, \label{E:PM1}\\
	\frac{d v_p^{+}}{dt^{+}} &= \left(\frac{\rho_f}{\rho_p}-1\right)\frac{1}{Fr^{2}} 
		+ \frac{\rho_f A_p}{2m_p} c_d \cdot \left| u^{+}-v_p^{+} \right| \cdot (u^{+}-v_p^{+})
		+ \frac{1}{{U_{\infty}^{2}}{L_{\infty}^{2}}} \frac{F_a^{+}}{m_p}, \label{E:PM2} \\
	\frac{d\omega_p^{+}}{dt^{+}} &= \frac{\rho_f {r_p^{+}}^5}{2 I_p}\, c_R \cdot \left| \frac{1}{2}\nabla^{+}\times u^{+} -\omega_p^{+}\right| \cdot \left(\frac{1}{2}\nabla^{+}\times u^{+} -\omega_p^{+}\right)\;. \label{E:PR3}
\end{align}
\end{subequations}
Here $Fr = \sqrt{U_\infty^2/(L_\infty g)}$ denotes the Froude number. All the non--dimensional quantities are represented by '$+$'.
\begin{table}[h!]
\caption{Typical values for the used parameters.}
\label{tab:1}
\begin{tabular}{lrrl}
\hline
Parameter	& Value	& Unit	& Reference \\ \hline
$\rho_f$ 	& $1000$	& $\mathrm{kg}\,\mathrm{m}^{-3}$	& water at $20^\circ$ C\\
$\nu$		& $10^{-6}$	& $\mathrm{m}^2\, \mathrm{s}^{-1}$	& water at $20^\circ$ C\\ 
$U_\infty$	& $ 10^{-2}$	& $\mathrm{m}\, \mathrm{s}^{-1}$	& considered geometry\\
$L_\infty$	& $ 10^{-1}$	& $\mathrm{m}$			& considered geometry\\
$T_{\infty}$ & $10$ & $\mathrm{s}$	& \\
$n$			& $\approx 2500$ &								& typical number of particles \\
$r_p$		& $0.5$--$2500\cdot 10^{-6}$ & $\mathrm{m}$		& microplastic particles \\ 
$\rho_p$	& $960$--$1140$	& $\mathrm{kg}\,\mathrm{m}^{-3}$	& e.g. polyamid, polyester or polystyrene \\
$A_p$		& $\pi r_p^2$ 	& $\mathrm{m}^2$				& spherical particle geometry \\
$m_p$		& $\frac{4}{3}\pi r_p^3 \rho_p$ & $\mathrm{kg}$	& spherical particle geometry \\
$I_p$		& $\frac{2}{5} m_p r_p^2$ & $\mathrm{kg}\,\mathrm{m}^2$	& spherical particle geometry \\
$K$			& $1$--$100\cdot 10^{-8}$ & $\mathrm{kg}\,\mathrm{m}\,\mathrm{s}^{-2}$\\
$\delta$	& $0.1\cdot L_\infty$		& $\mathrm{m}$ \\
$\epsilon$	& $10^{-2}\cdot L_\infty$	& $\mathrm{m}$ \\
$Re$		& $1000$ \\
${Fr}^{2}$ 		& $1.02\cdot 10^{-4}$\\
\hline
\end{tabular}
\end{table} 

\section {Numerical Methods}
In our model we assume only a one--way coupling of the flow field to the particles; the back--coupling from the particles on the flow is neglected. Hence we are able to split the numerical computations: First we solve the Navier--Stokes equations~(\ref{E:NS1}, \ref{E:NS2}) in the cavity with the obstacle but excluding the particles. The obtained stationary flow field is used to integrate the transient equations of motion of the particles~(\ref{E:PM1}, \ref{E:PM2}, \ref{E:PR3}) to determine their trajectories. 
\par To simulate the stationary Navier--Stokes equations~(\ref{E:NS1}, \ref{E:NS2}), the computational domain is discretized using a staggered grid with square cells. The pressure variable is located in the cell center, the horizontal and vertical velocities are located at the midpoints of the right or top cell edge. Chorin's projection is used to treat the incompressibility condition~\ref{E:NS1}, see~(\cite{chorin1968numerical, griebel1997numerical, seibold2008compact, strang2007computational}) for details. The diffusive term and the resulting pressure Laplacian are implicitly solved using a LU--decomposition of the linear systems. A virtual time stepping is applied until stationary is reached. 

\subsection{Code validation}
\begin{figure}[h!]
  \includegraphics[width=.45\linewidth]{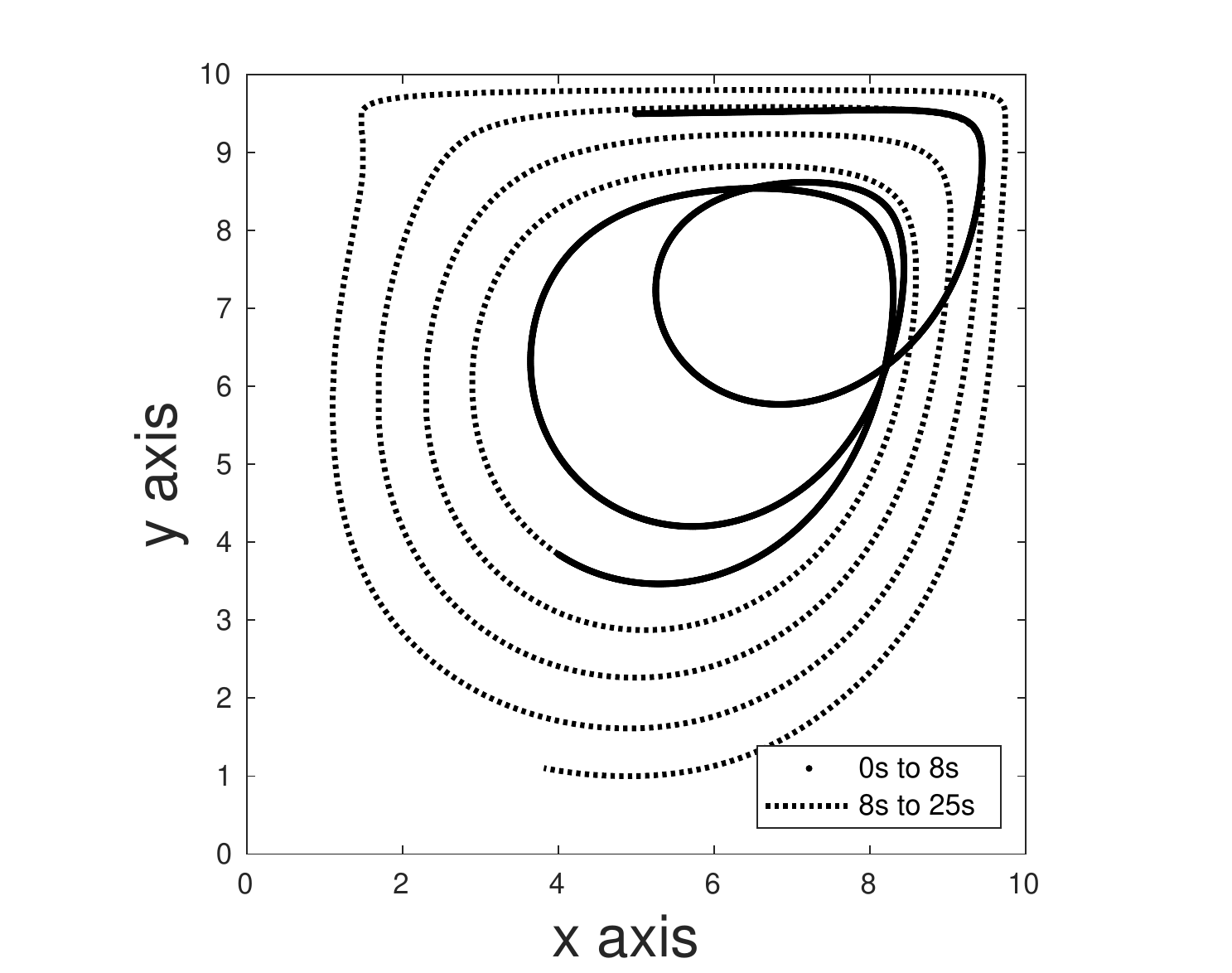}
  \hfill
  \includegraphics[width=.35\linewidth]{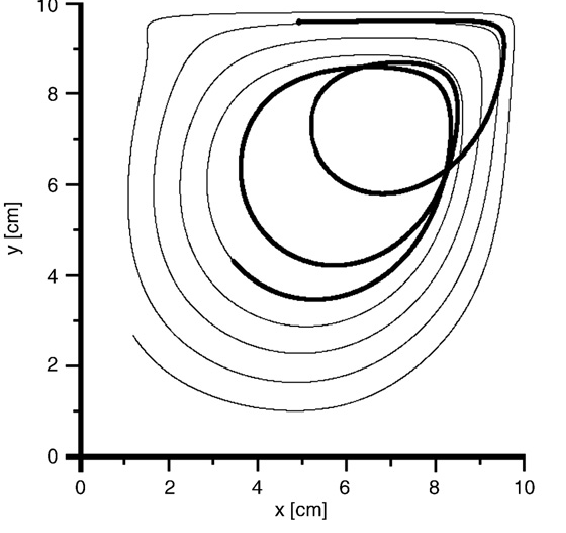}
\caption{Comparison of our simulation result (left) to the the result introduced by \cite{kosinski2009simulation} (right; with kind permission to reuse from the Elsevier copyright clearance center).}
\label{fig:2}
\end{figure}

Various works based on the behavior of solid particle in a lid-driven cavity have been proposed by several researchers. In order to check the validation of our code, we compared the present simulation result with the result reported by \cite{kosinski2009simulation}. Please note that, in this case, all the necessary parameters and equations are considered according to their assumption. For the reference model, a lid-driven cavity with 0.1 [$\mathrm{m}$] long side and a constant velocity of 0.175 [$\mathrm{m}$ $\mathrm{s}^{-1}$] on the top moving boundary is considered. The kinematic viscosity of the fluid is assumed to be 3.72 $\times10^{-5}[\mathrm{m}^2\, \mathrm{s}^{-1}]$. The Reynolds number of the fluid is derived from the above data, and it is  Re= 470. Figure \ref{fig:2} shows a comparison of the present simulation result (left) and results due to \cite{kosinski2009simulation} (right).

\subsection{Fluid streamlines}
\begin{figure}[h!]
\begin{minipage}{.48\textwidth}
  \includegraphics[width=\linewidth]{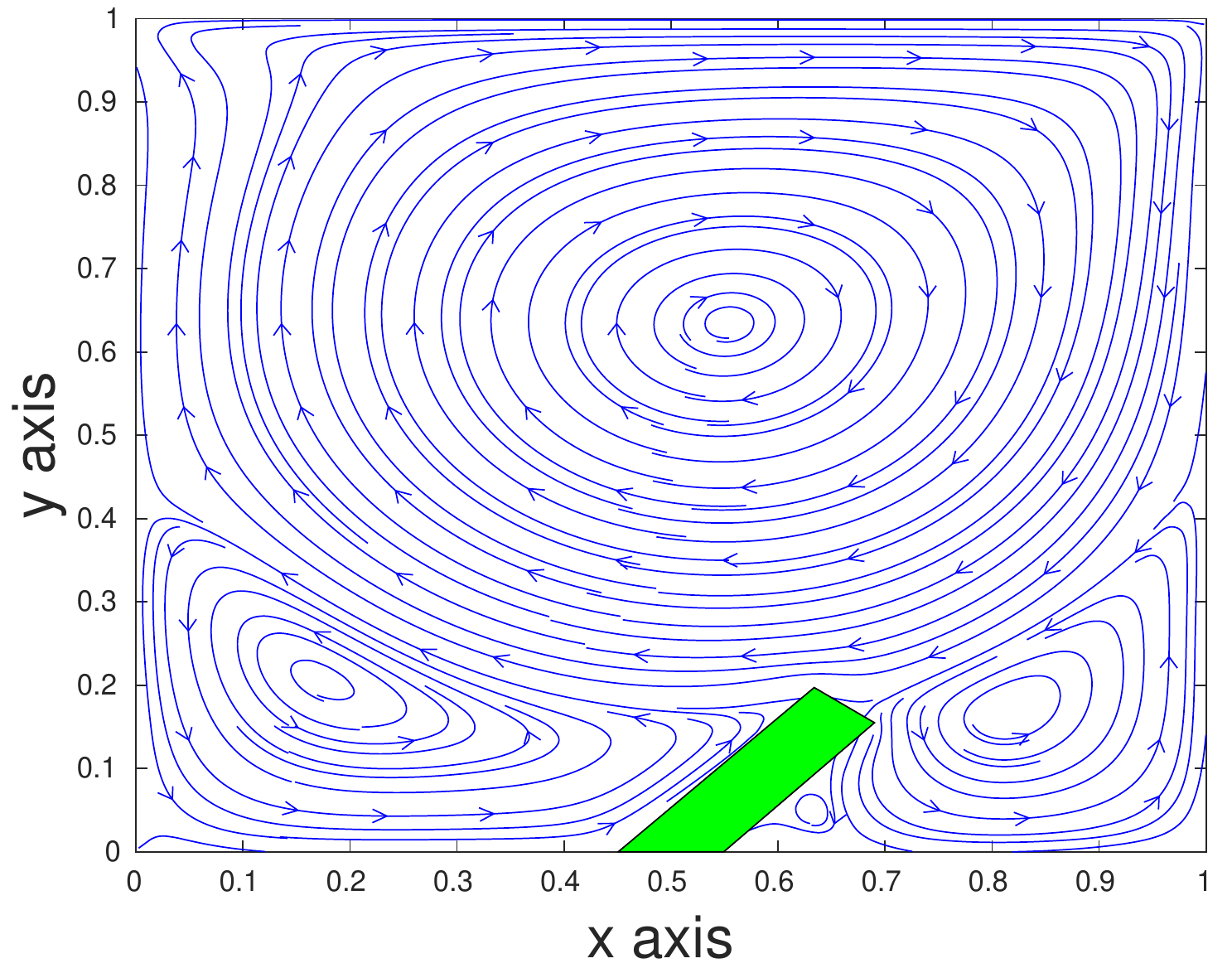}
  
  \centerline{(a)}
  \end{minipage}
  \hfill 
  \begin{minipage}{.48\textwidth}
  \includegraphics[width=\linewidth]{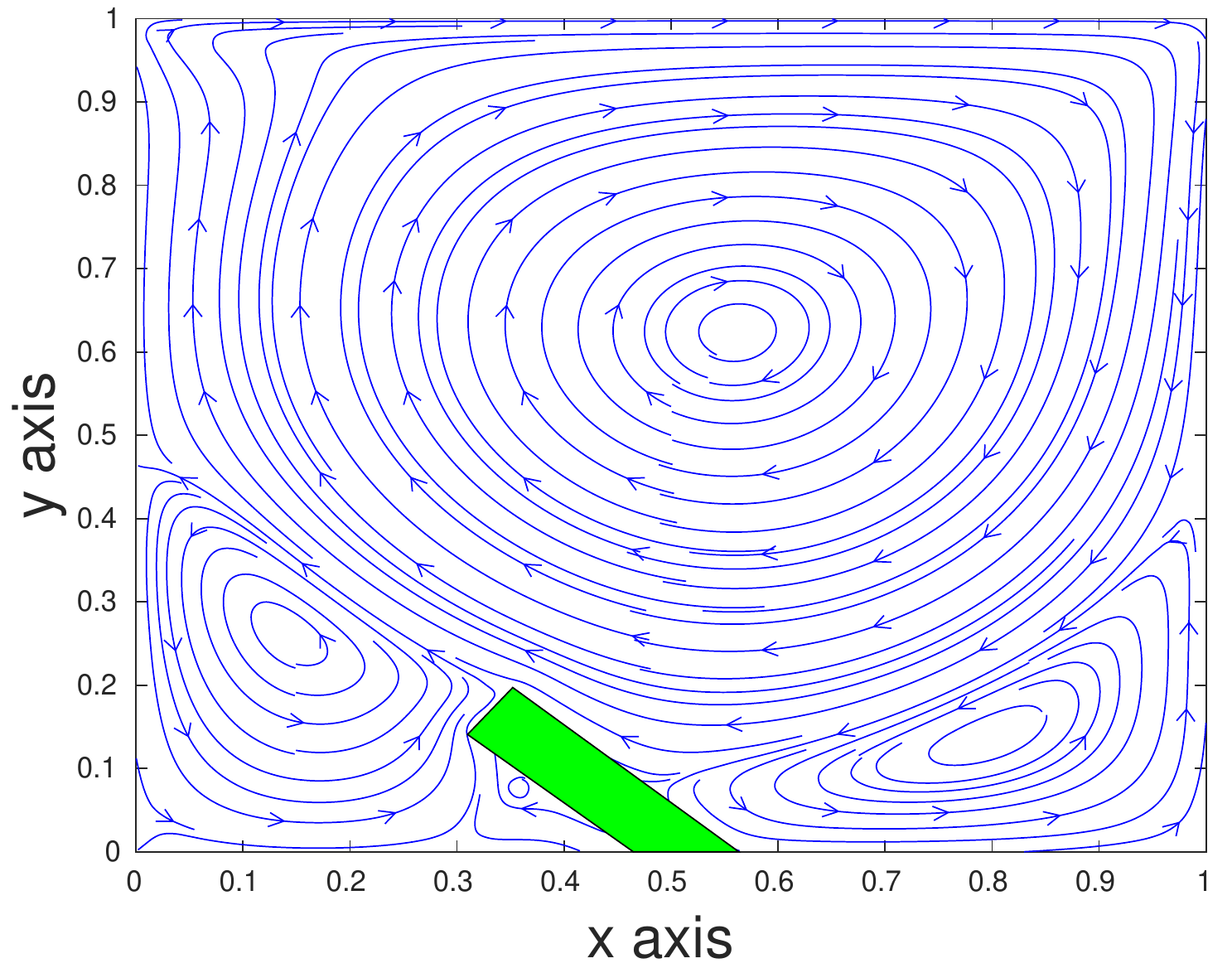}

  \centerline{(b)}
  \end{minipage}

\begin{minipage}{.48\textwidth} 
  \includegraphics[width=\linewidth]{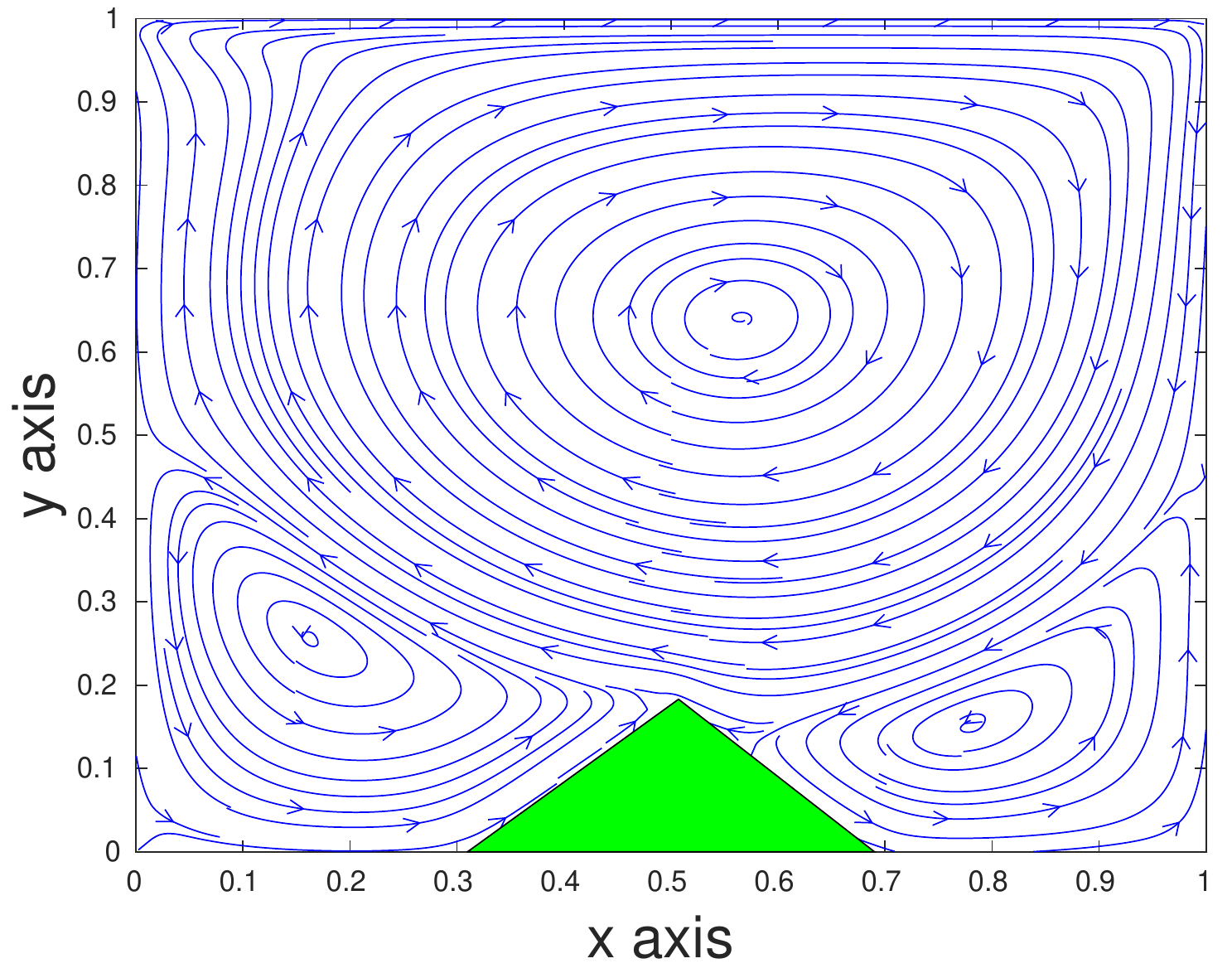}

  \centerline{(c)}
  \end{minipage}
  \hfill 
  \begin{minipage}{.48\textwidth}
  \end{minipage}
 \caption{Fluid streamlines at $Re=1000$ for three types of obstacles: Acute angled obstacle (a), obtuse angled obstacle (b) and triangular obstacle (c).}
 \label{fig:3}
 \end{figure}

The fluid streamlines for Reynolds number $Re=1000$, with different obstacles in the lid--driven cavity, are shown in Figure~\ref{fig:3}. The obstacles are shown in green color and are labeled as Acute Angled Obstacle--AAO (see Figure.~\ref{fig:3} (a)), Obtuse Angled Obstacle--OAO (see Figure.~\ref{fig:3} (b)) and Triangular Obstacle--TGO  (see Figure.~\ref{fig:3} (c)) respectively. In all these figures, we observe a primary vortex appearing in the center of the cavity and two secondary vorticity in the two bottom corners of the cavity. Figures \ref{fig:3} (a) and \ref{fig:3} (b) indicate even the presence of a third secondary vortex of low intensity in the region below the obstacle. 

In order to obtain the particle trajectories, the equations of motion~(\ref{E:PM1}, \ref{E:PM2}, \ref{E:PR3}) are integrated using the previously computed flow field $U$. The numerical scheme is based on the Euler method combined with a linear interpolation of the flow velocity $U$ in--between the grid points used for the Navier--Stokes computation. Typically, we used any rather coarse grid for the flow computations ($71\times 71$ grid) and a time step of $\Delta t=2\times 10^{-4}$ for the particle trajectories.

Several studies suggest to consider the particle collision with the cavity wall; due to the migration of particles towards the wall (\cite{kosinski2009simulation,safdari2014lattice}). Hence, In this study, we considered hard particle collision with the cavity wall as reported in (\cite{schwarzkopf2011multiphase,kartushinskii2004numerical}). The restitution coefficient is considered as 0.9. However, the particle-obstacle wall collision is not considered; due to the assumption that the particles wedge to the obstacle after touch.

\section{Results and Discussion}

\subsection{Particle trajectories without attachment force}
\begin{figure}[h!]
\begin{minipage}{.48\textwidth}
  \includegraphics[width=\linewidth]{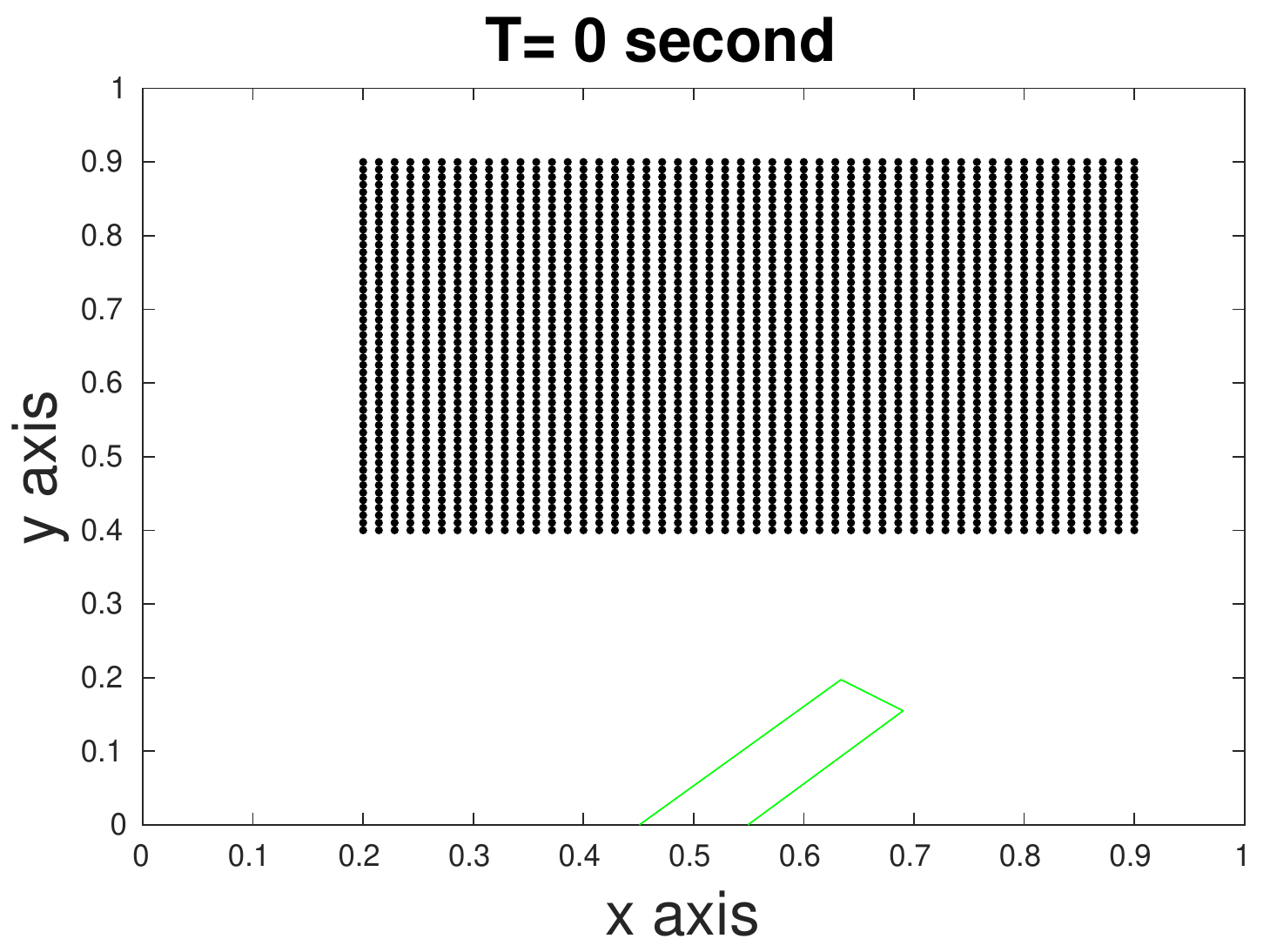}

\centerline{(a)}
\end{minipage}
\hfill
\begin{minipage}{.48\textwidth}
  \includegraphics[width=\linewidth]{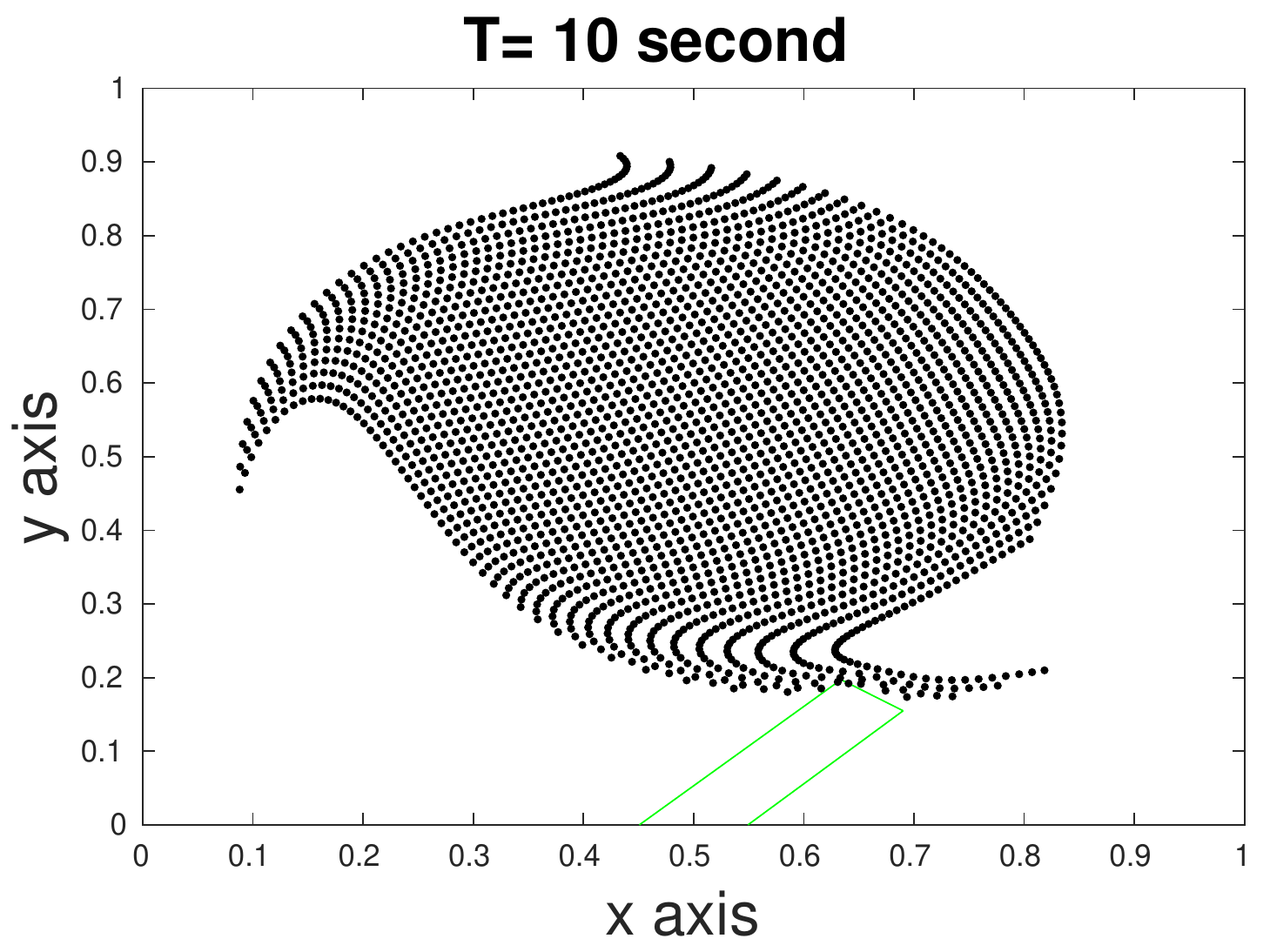}

\centerline{(b)}
\end{minipage}

\begin{minipage}{.48\textwidth}
  \includegraphics[width=\linewidth]{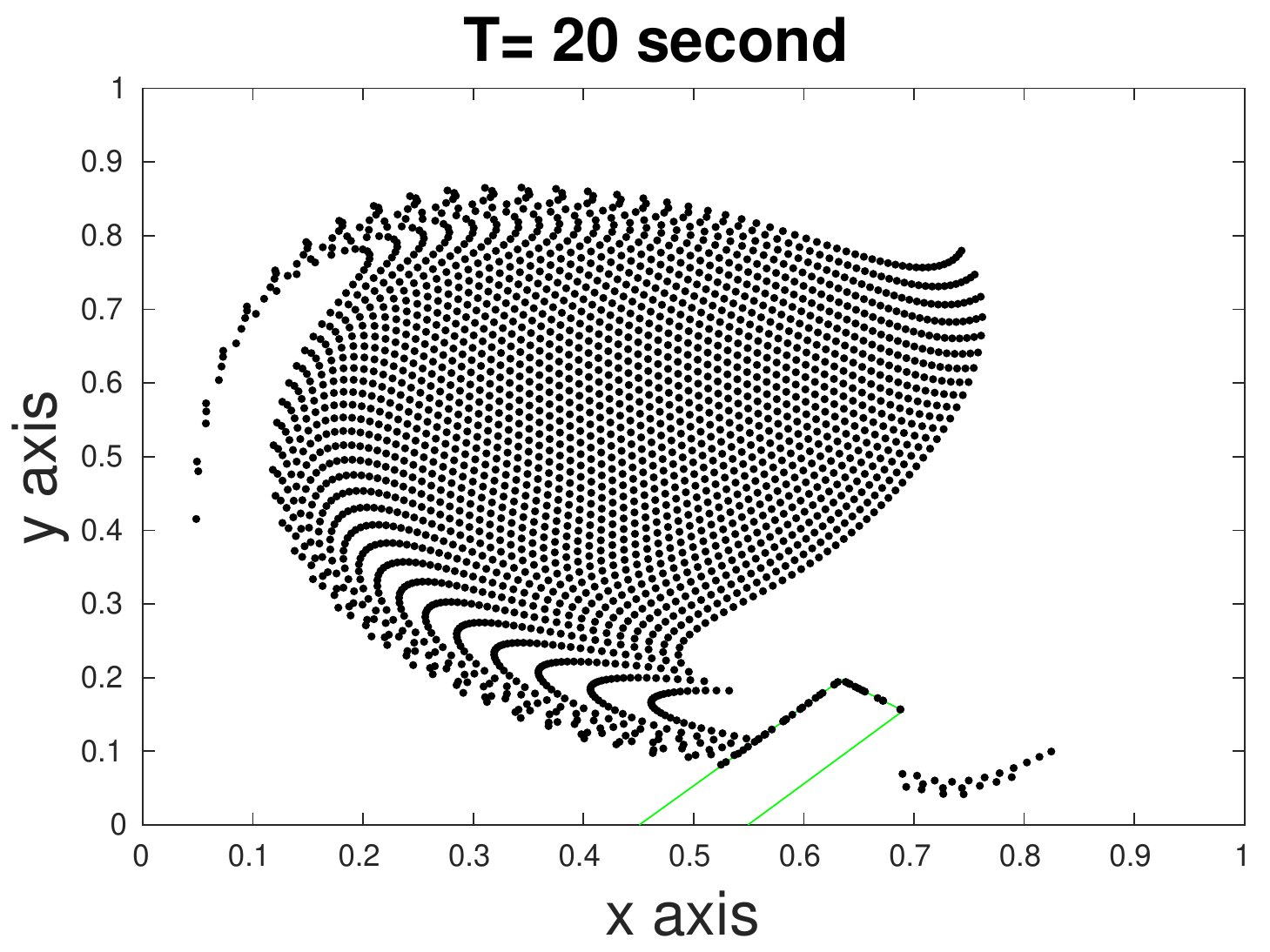}

\centerline{(c)}
\end{minipage}
\hfill
\begin{minipage}{.48\textwidth}
  \includegraphics[width=\linewidth]{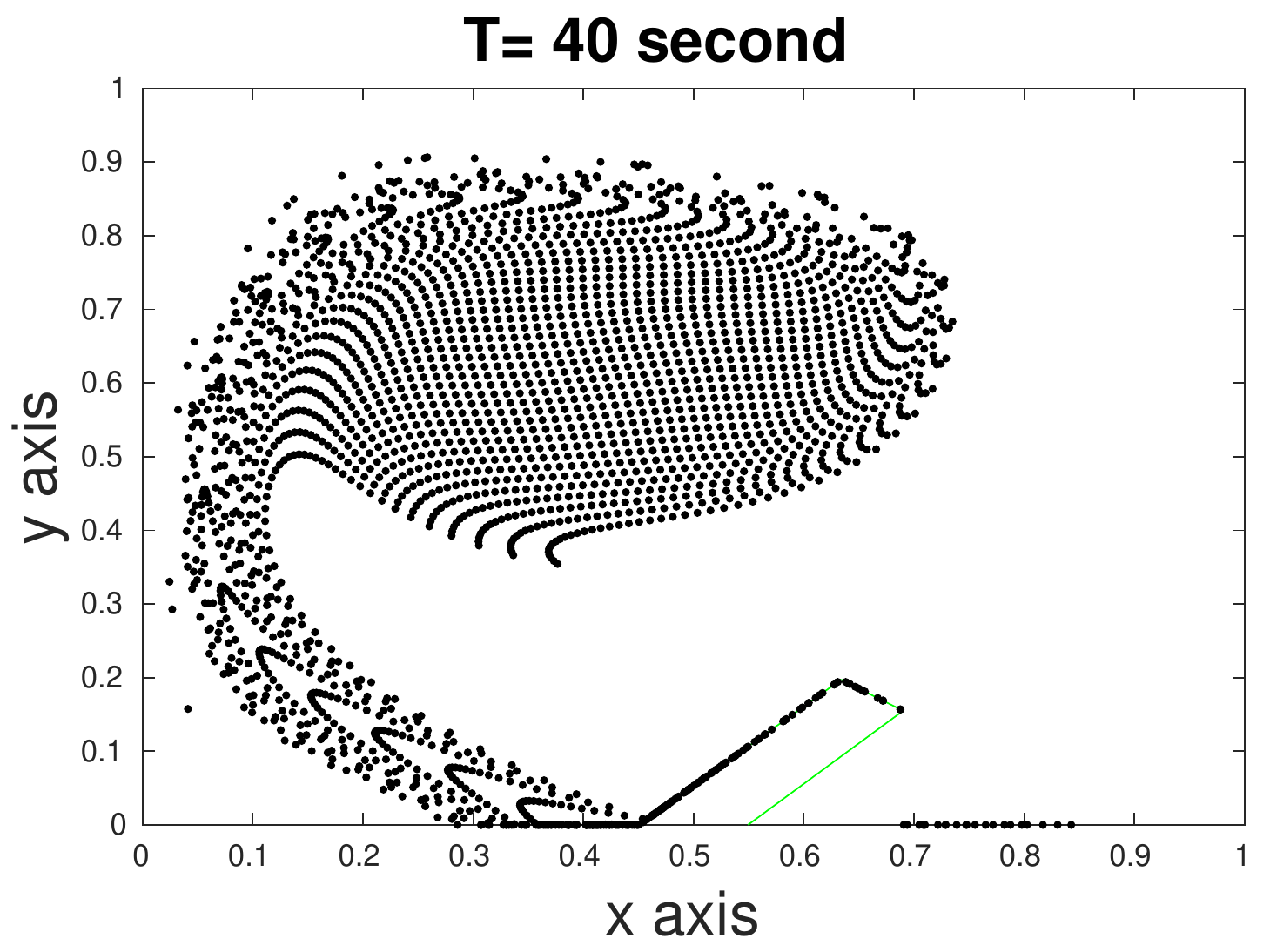}

\centerline{(d)}
\end{minipage}

\begin{minipage}{.48\textwidth}
  \includegraphics[width=\linewidth]{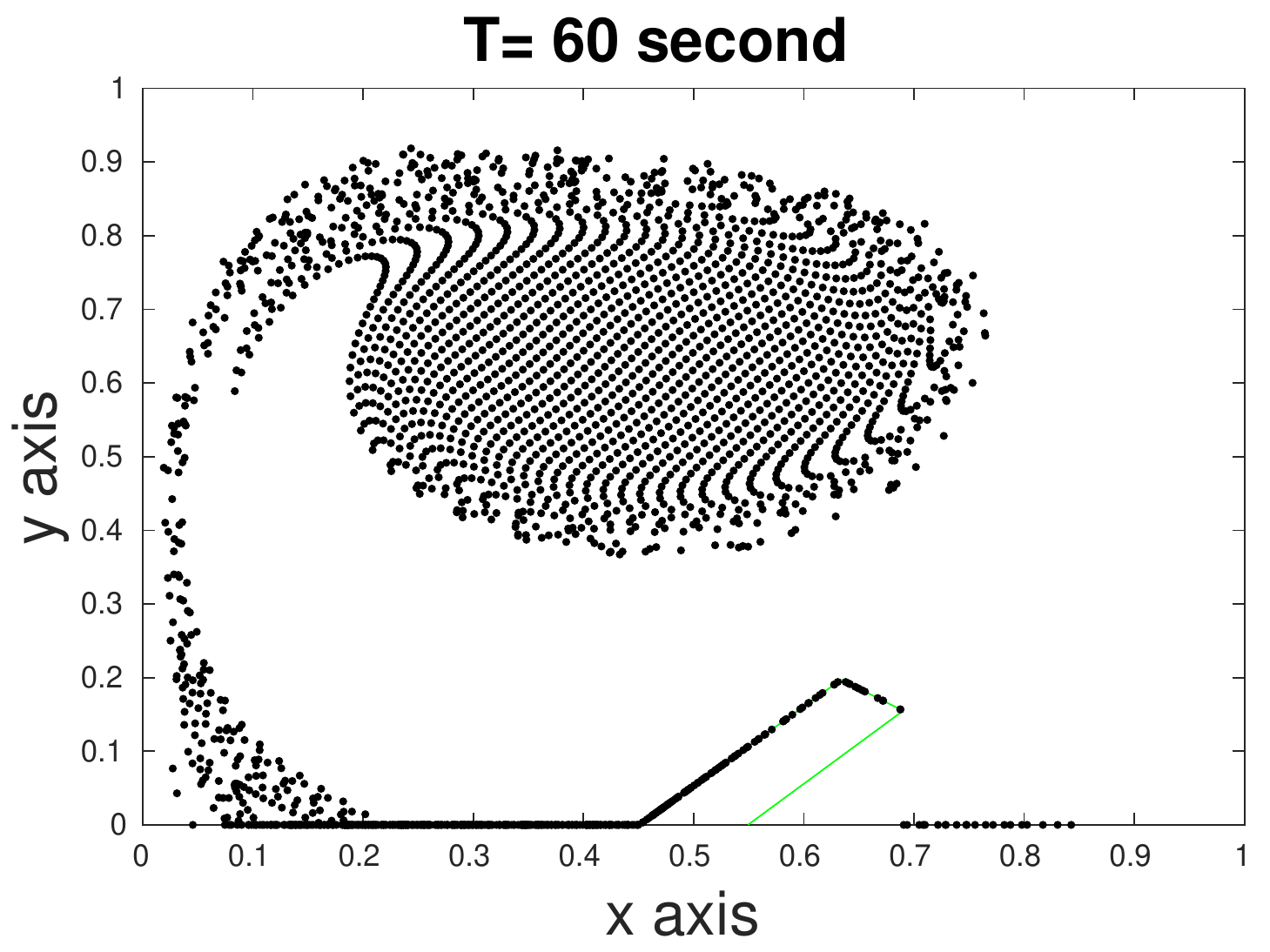}

\centerline{(e)}
\end{minipage}
\hfill
\begin{minipage}{.48\textwidth}
  \includegraphics[width=\linewidth]{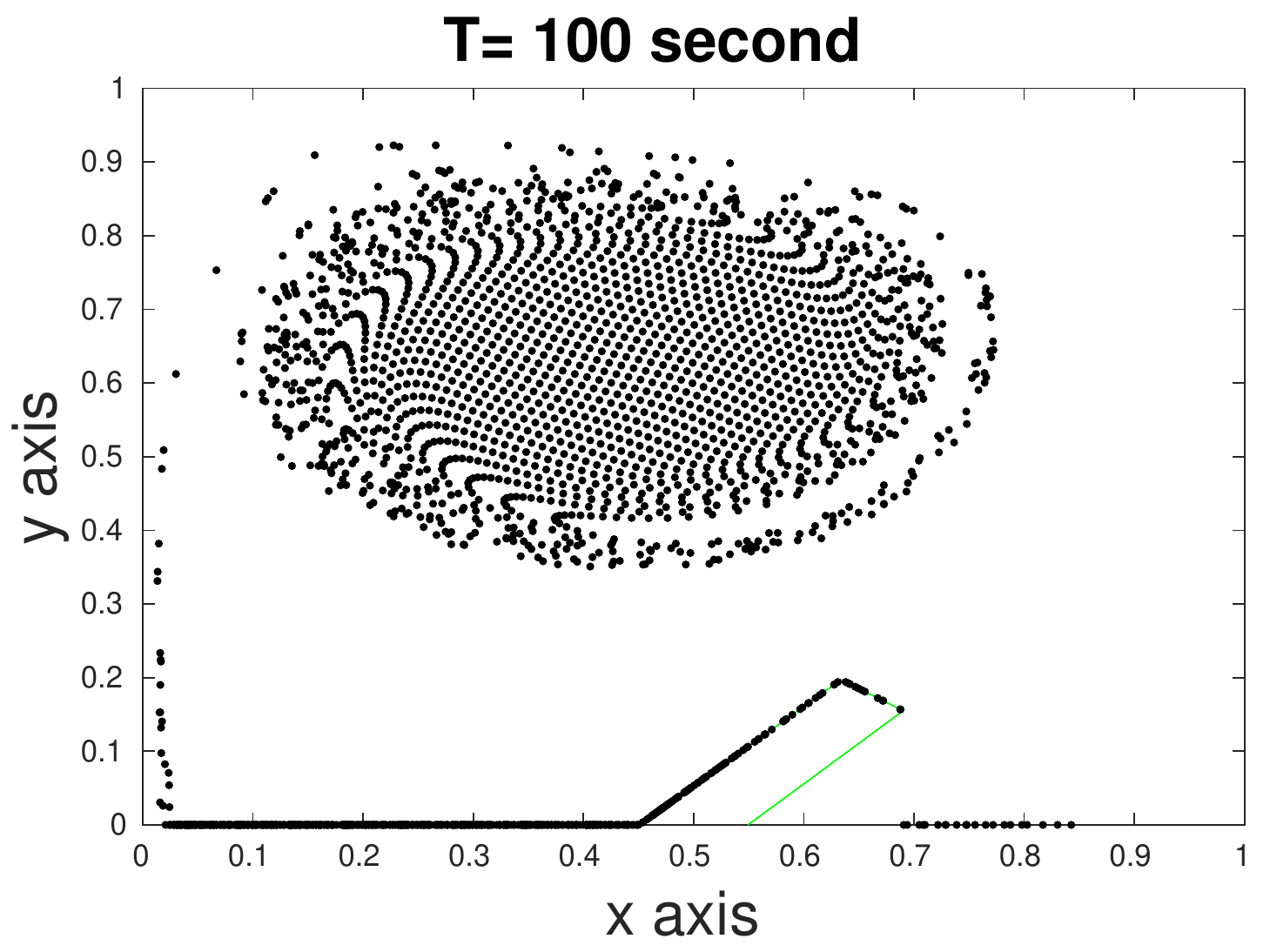}

\centerline{(f)}
\end{minipage}

\caption{Particles position at different times without attachment force on the AAO.}
\label{fig:4}
\end{figure}

  The movement of particles at different times without the attachment force in the lid-driven cavity in the presence of AAO is shown in Figure \ref{fig:4}. In total, 2500 particles are considered for the simulation. The size of the particles is considered to be the same, which is $1\times10^{-4}$[m]. Initially, all particles are placed uniformly inside the rectangular domain $[0.2, 0.9]\times[0.4, 0.9]$, instead of the whole domain, to avoid the particle collision with the wall. 
It is noticed that the disturbance in the arrangement of the particles is started from the corner of the rectangular shape, and as time goes on, the rectangular arrangement of particles gets rearranged in circular shape. Due to the consideration of small size particles, it is observed that the particles flow with fluid streamline and construct trajectories almost similar to the fluid streamlines. The fluid velocity in the secondary vortices is less compared to the others part of the cavity, which leads to less influence of fluid on the particles. Therefore, the gravitational force dominates the movement of the particles inside the secondary vortices, thus results in particles settle after entering the secondary vortices. It can be seen that few particles are sedimented on the left side of the obstacle and some of the particles sedimented on the AAO. The number of sedimented particles on the AAO is shown in Table \ref{tab:4}.

\begin{figure}[h!]
\begin{minipage}{.48\textwidth}
  \includegraphics[width=\linewidth]{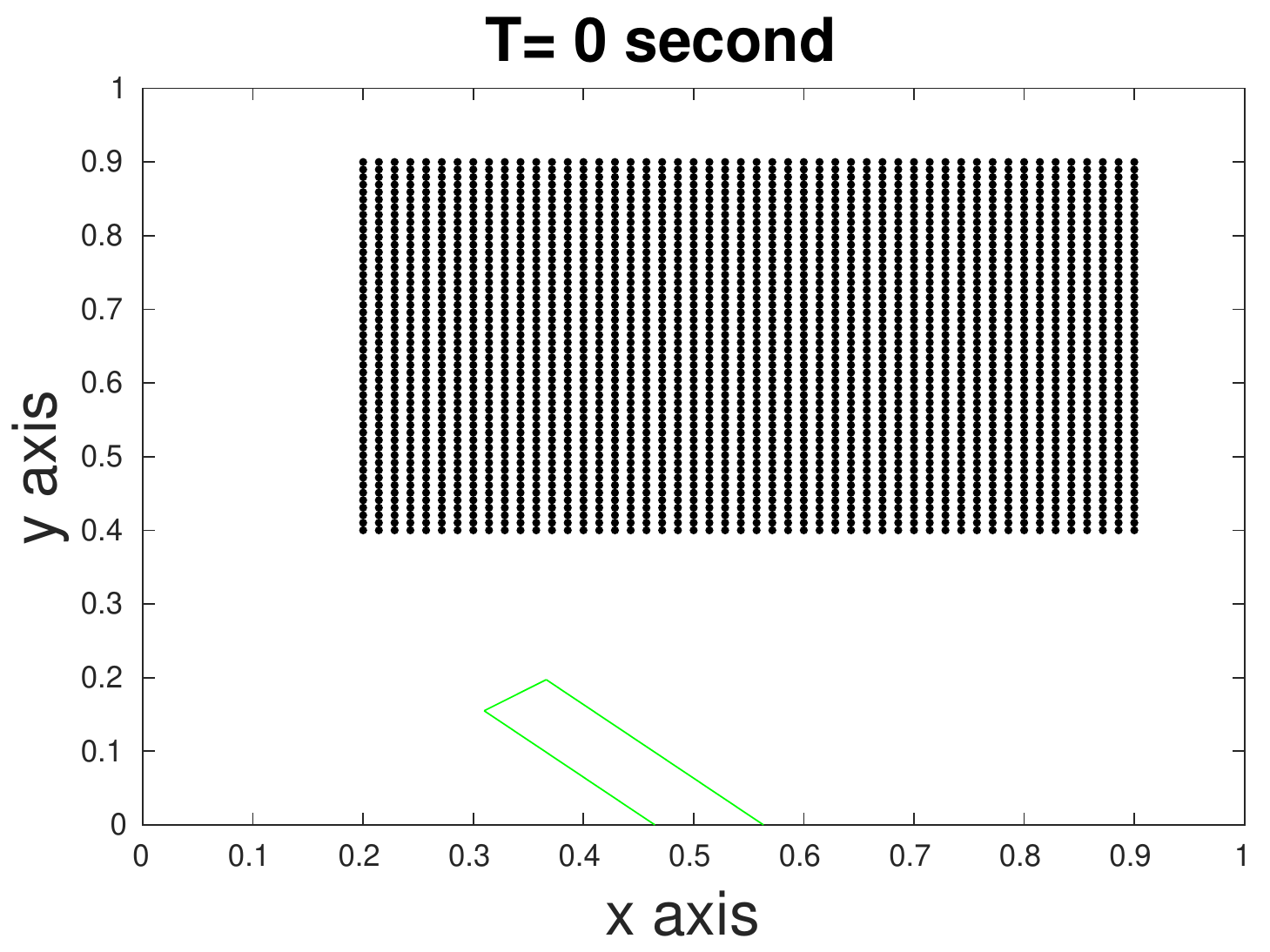}

\centerline{(a)}
\end{minipage}
\hfill
\begin{minipage}{.48\textwidth}
  \includegraphics[width=\linewidth]{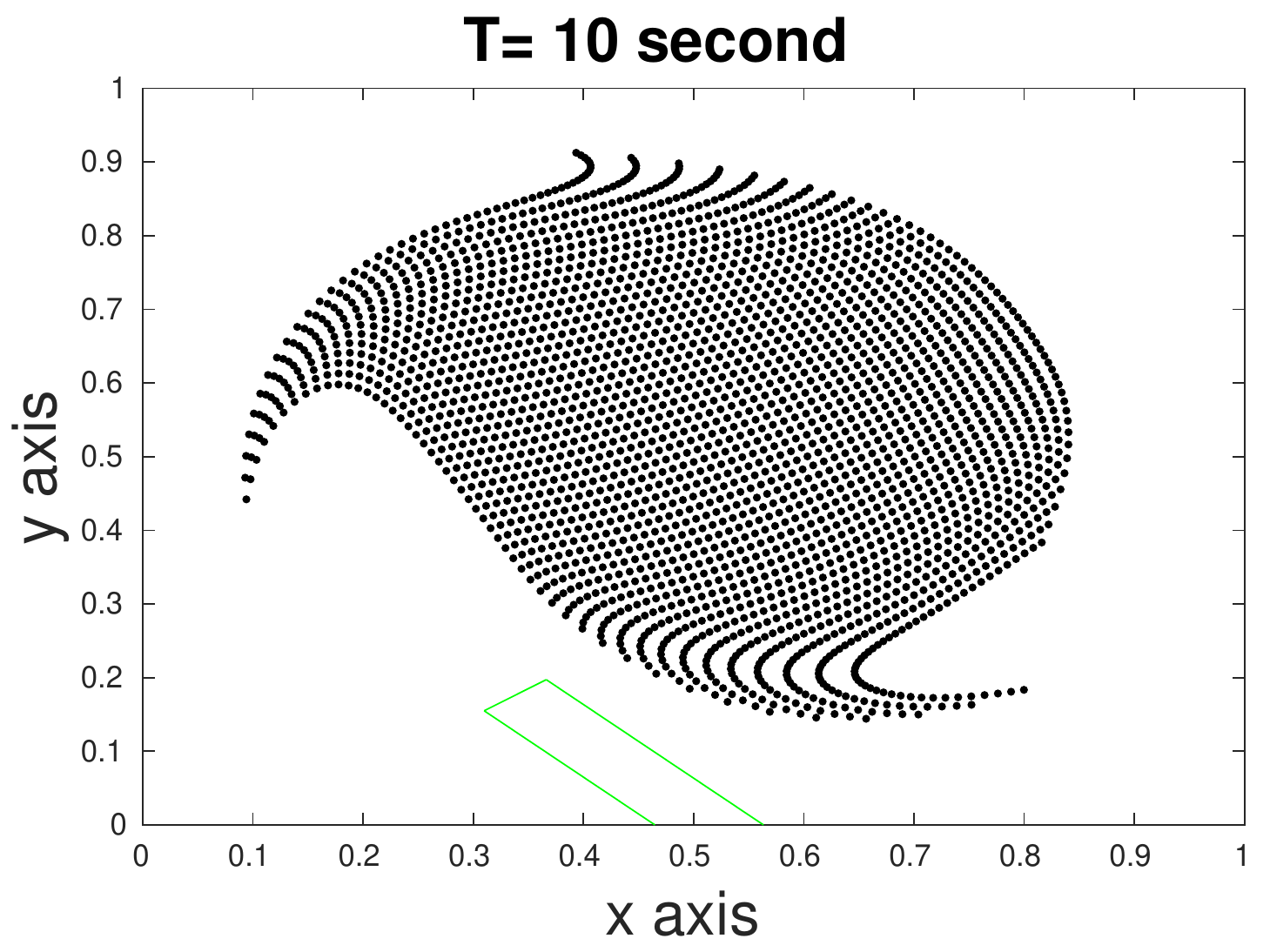}

\centerline{(b)}
\end{minipage}

\begin{minipage}{.48\textwidth}
  \includegraphics[width=\linewidth]{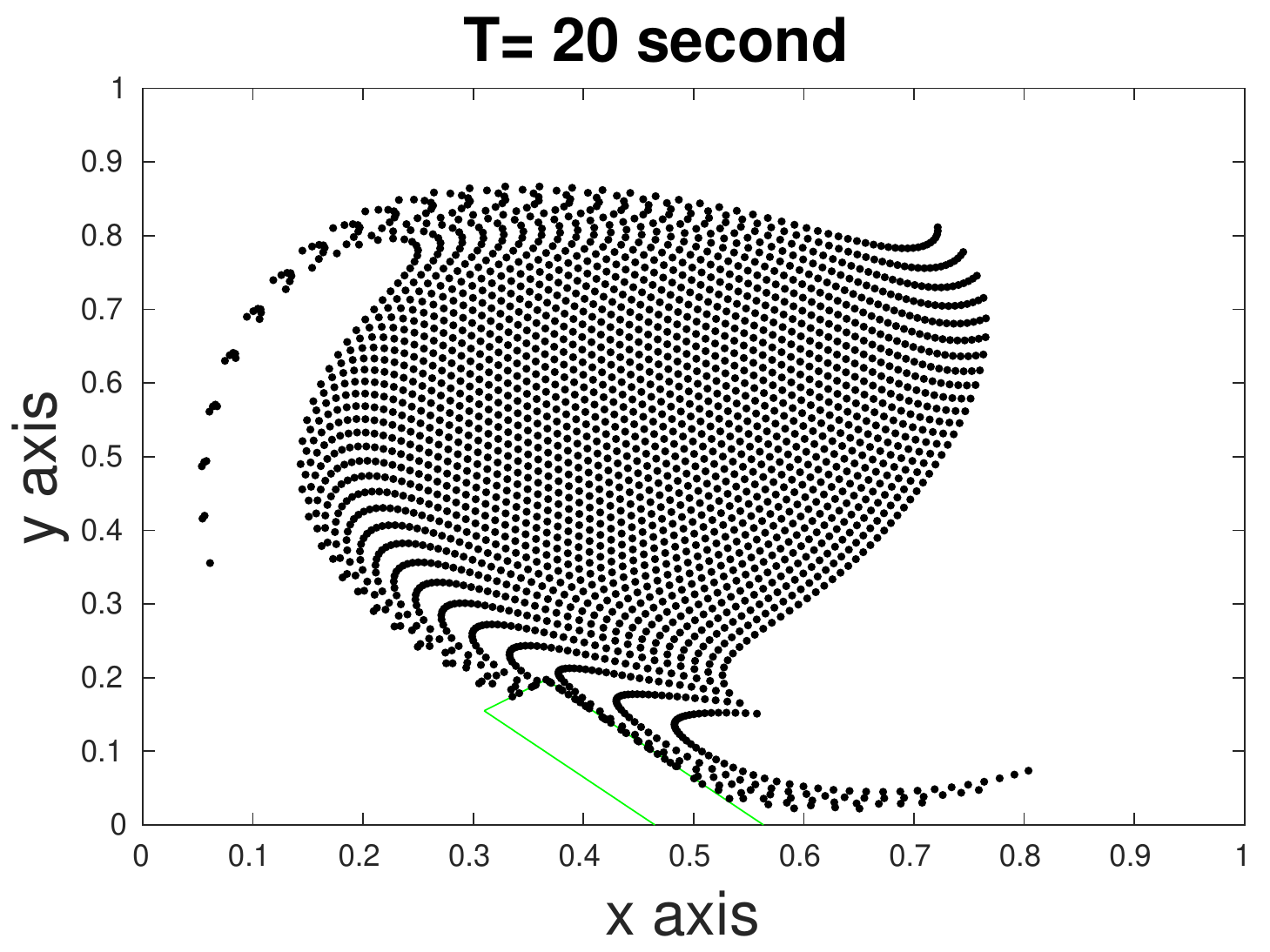}

\centerline{(c)}
\end{minipage}
\hfill
\begin{minipage}{.48\textwidth}
  \includegraphics[width=\linewidth]{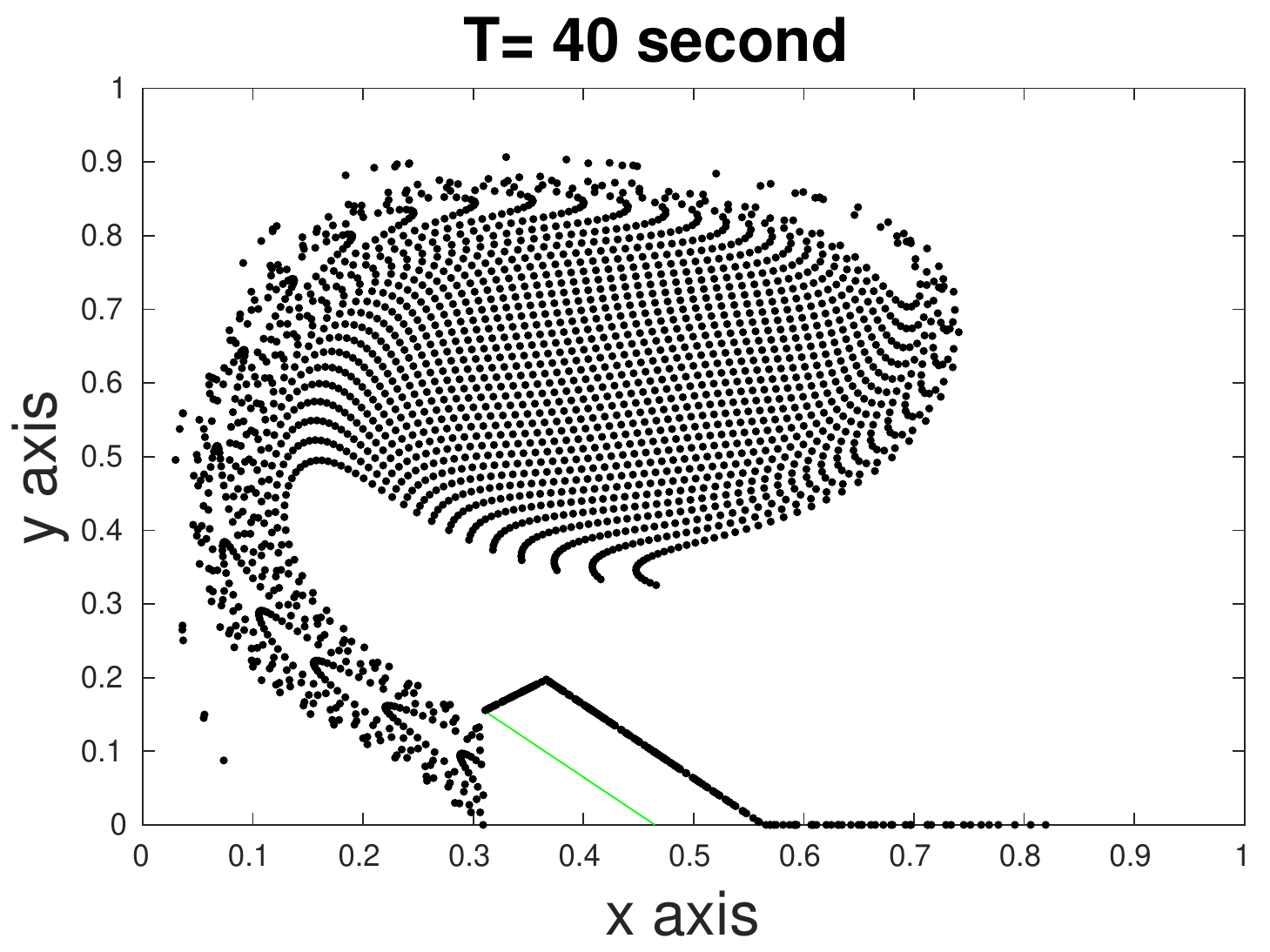}

\centerline{(d)}
\end{minipage}

\begin{minipage}{.48\textwidth}
  \includegraphics[width=\linewidth]{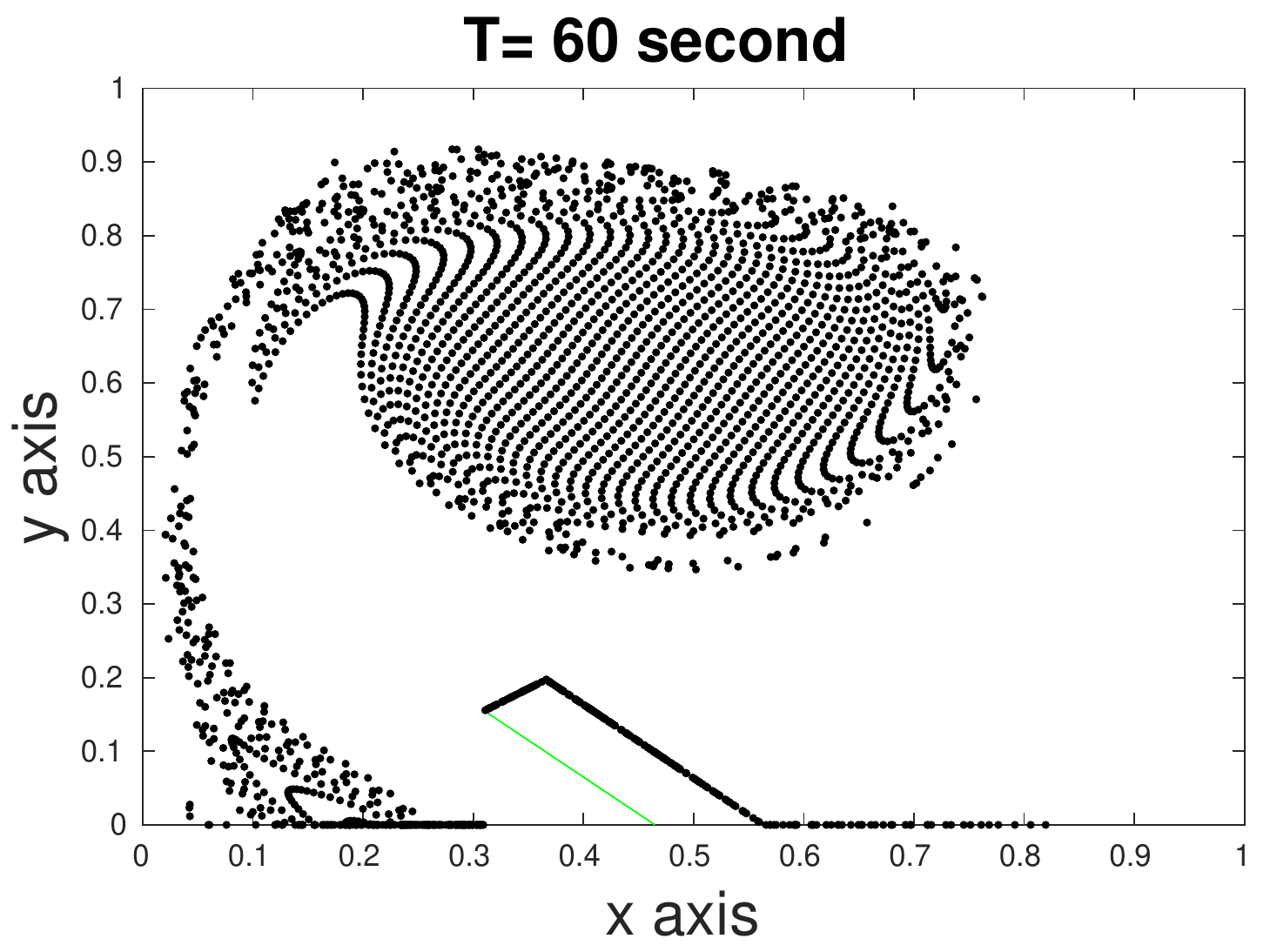}

\centerline{(e)}
\end{minipage}
\hfill
\begin{minipage}{.48\textwidth}
  \includegraphics[width=\linewidth]{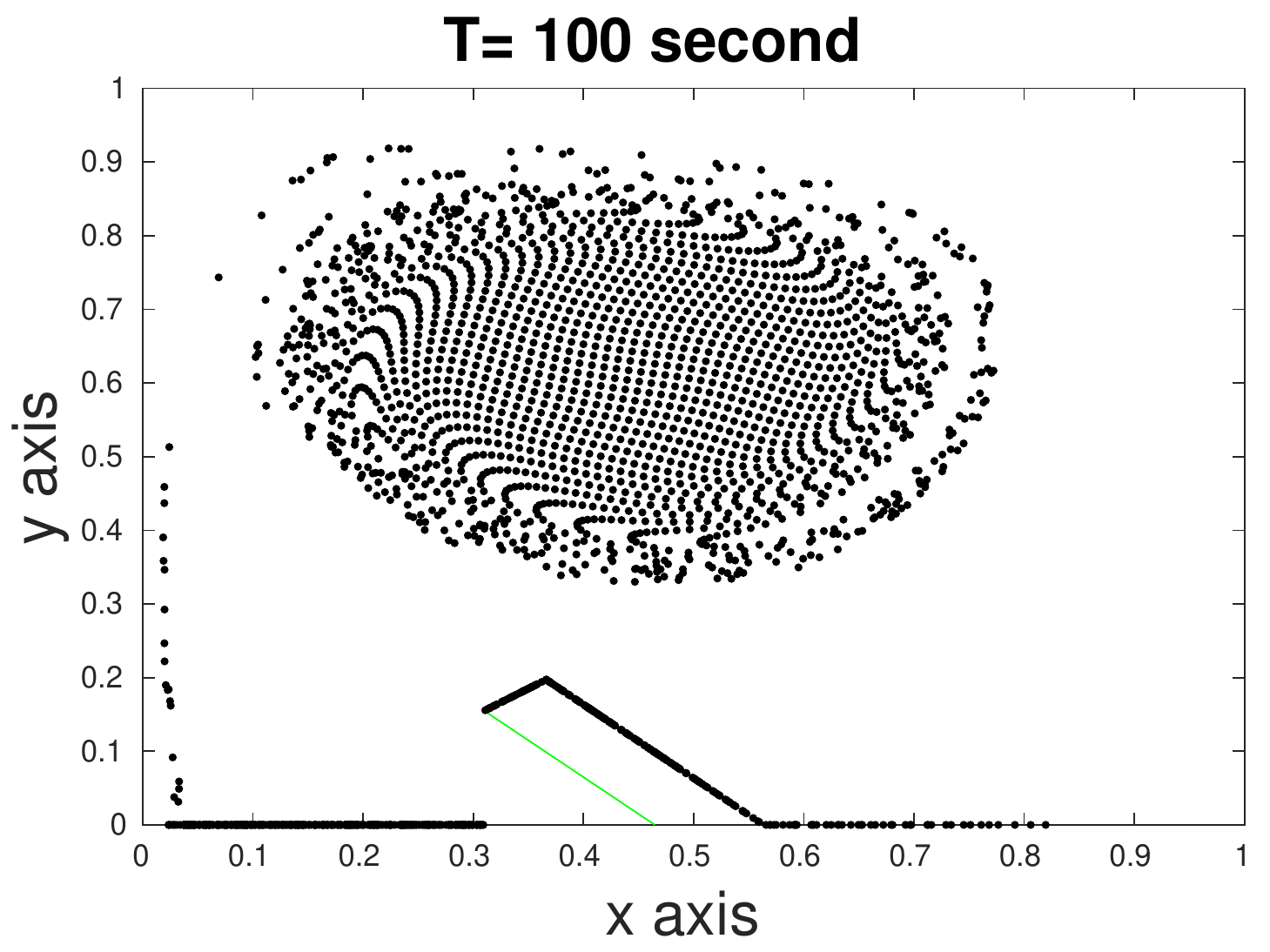}

\centerline{(f)}
\end{minipage}

\caption{Particles position at different times without attachment force on the OAO.}
\label{fig:5}
\end{figure}

Figure \ref{fig:5} shows the sedimentation of the particles on the OAO. The initial condition and all the other parameters are considered same as the simulation carried out for the AAO. In this case, the same trend has also been observed, i.e., as time goes on, the rectangular arrangement of the particles are rearranged as the circular shape. The more particles sedimented on the OAO than the AAO, which can be seen from Table \ref{tab:4}. The reason is that the OAO is inclined toward the left wall of the cavity. One can also notice that some particles sedimented on the left side of the obstacle. Therefore, the results show that the sedimentation also depends on the obstacle position.

\begin{figure}[h!]
\begin{minipage}{.48\textwidth}
  \includegraphics[width=\linewidth]{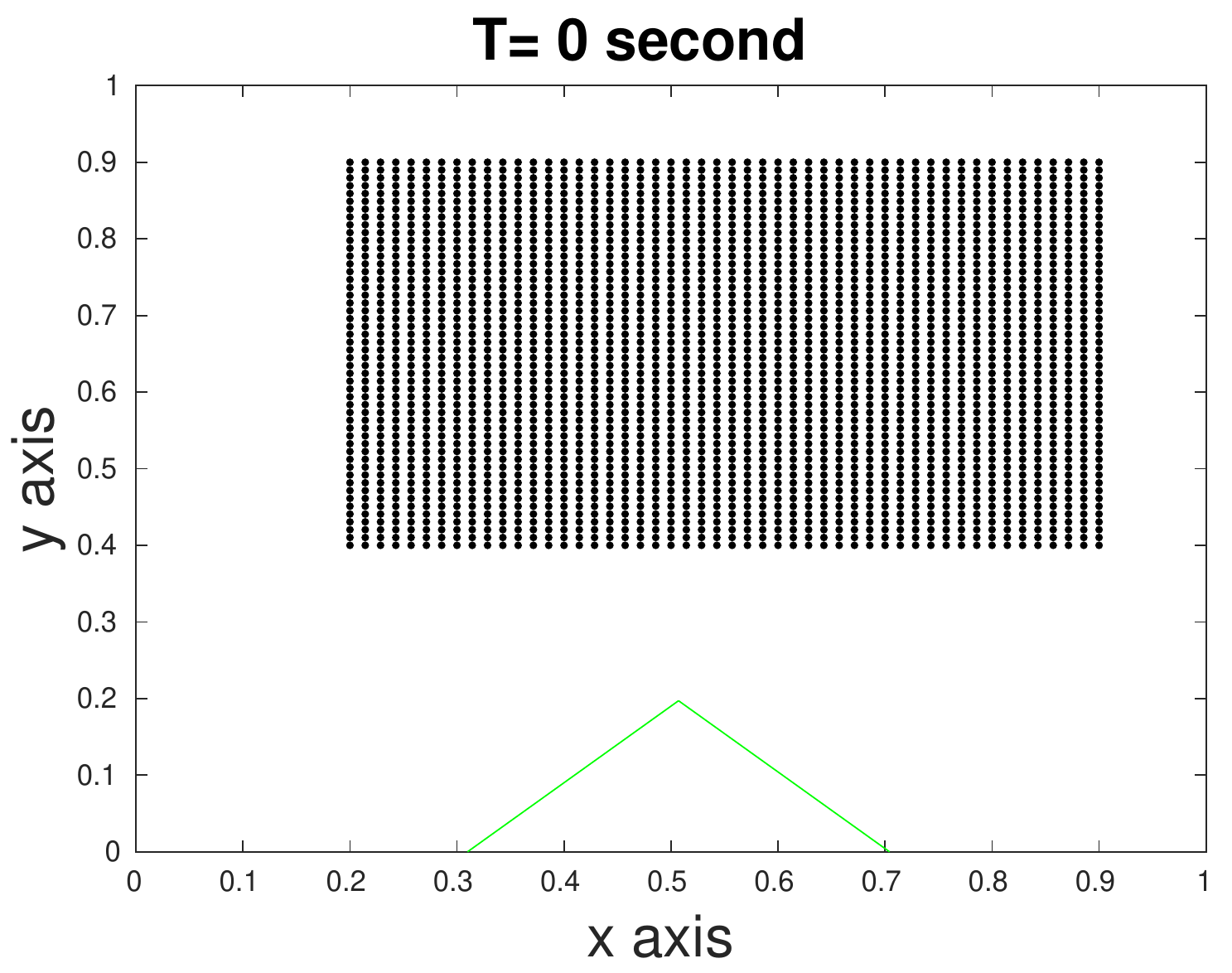}

\centerline{(a)}
\end{minipage}
\hfill
\begin{minipage}{.48\textwidth}
  \includegraphics[width=\linewidth]{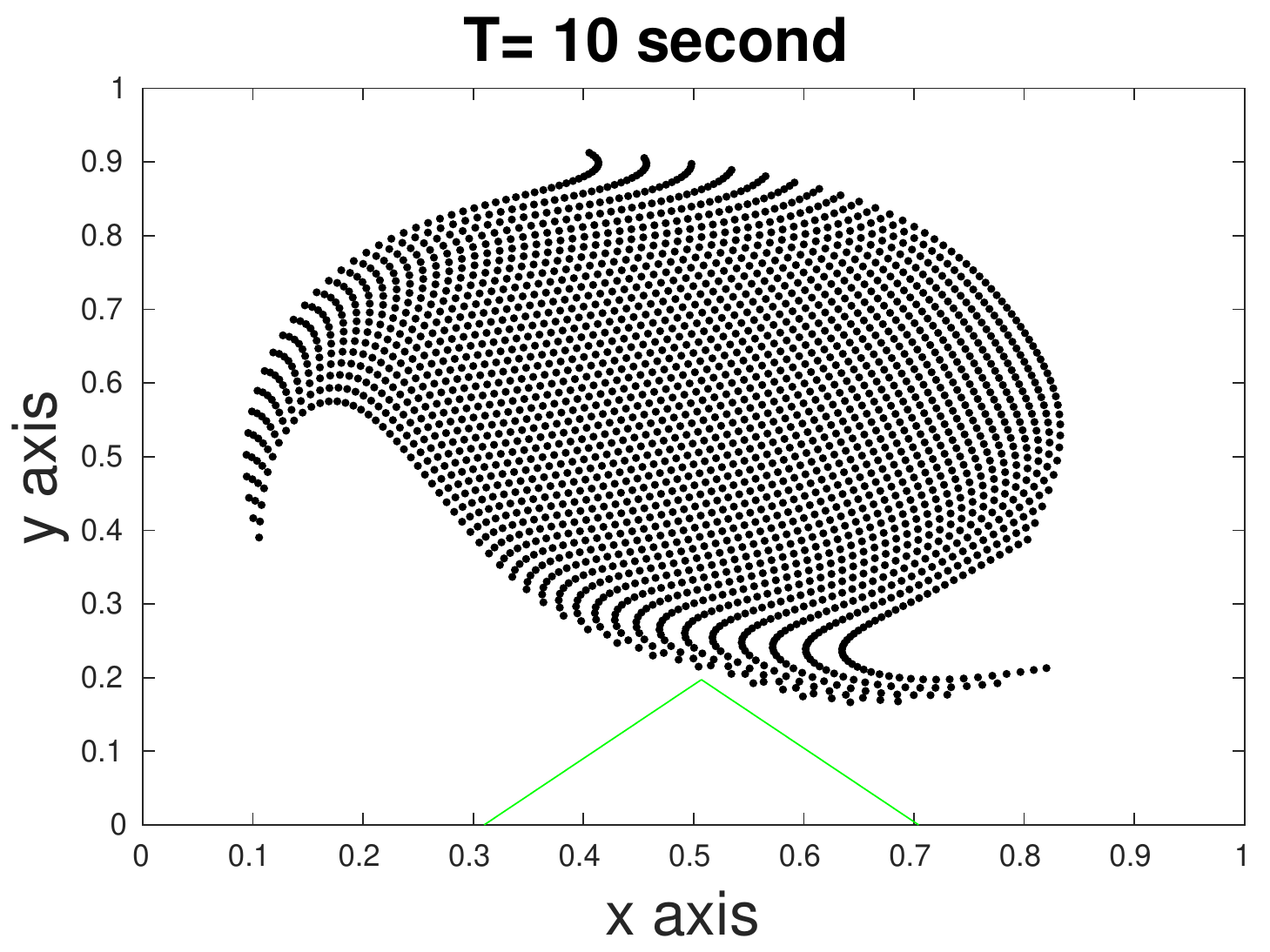}

\centerline{(b)}
\end{minipage}

\begin{minipage}{.48\textwidth}
  \includegraphics[width=\linewidth]{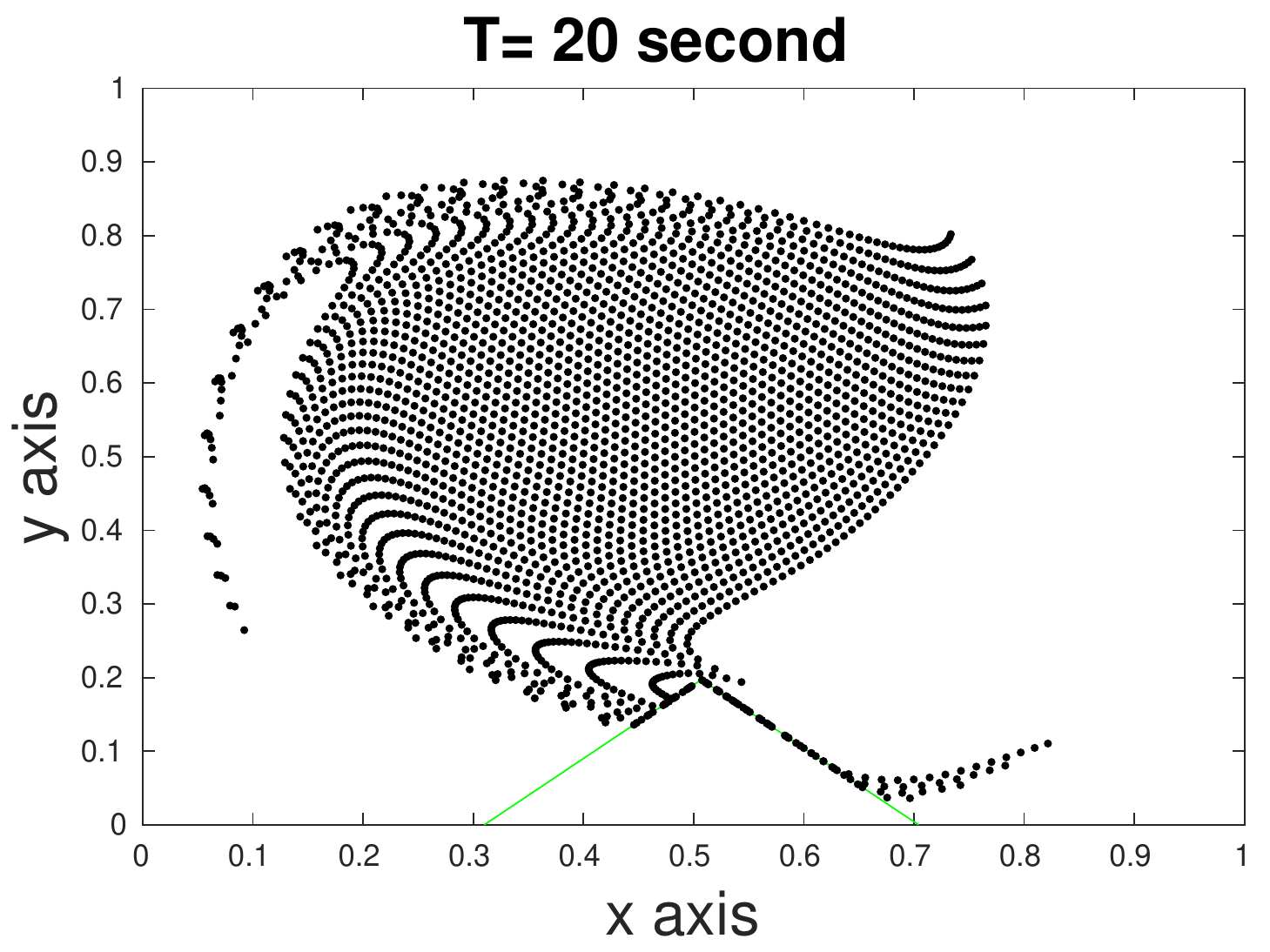}

\centerline{(c)}
\end{minipage}
\hfill
\begin{minipage}{.48\textwidth}
  \includegraphics[width=\linewidth]{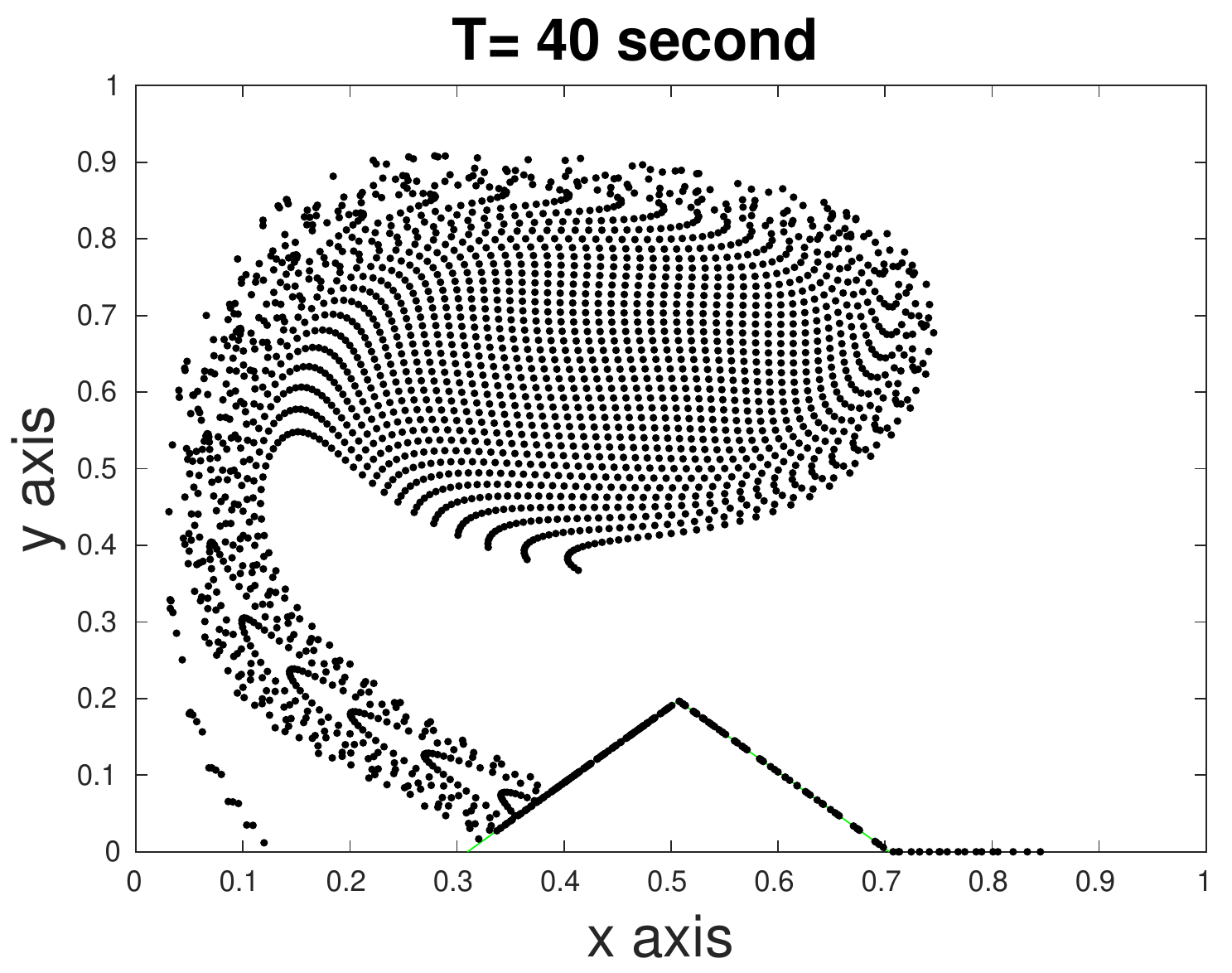}

\centerline{(d)}
\end{minipage}

\begin{minipage}{.48\textwidth}
  \includegraphics[width=\linewidth]{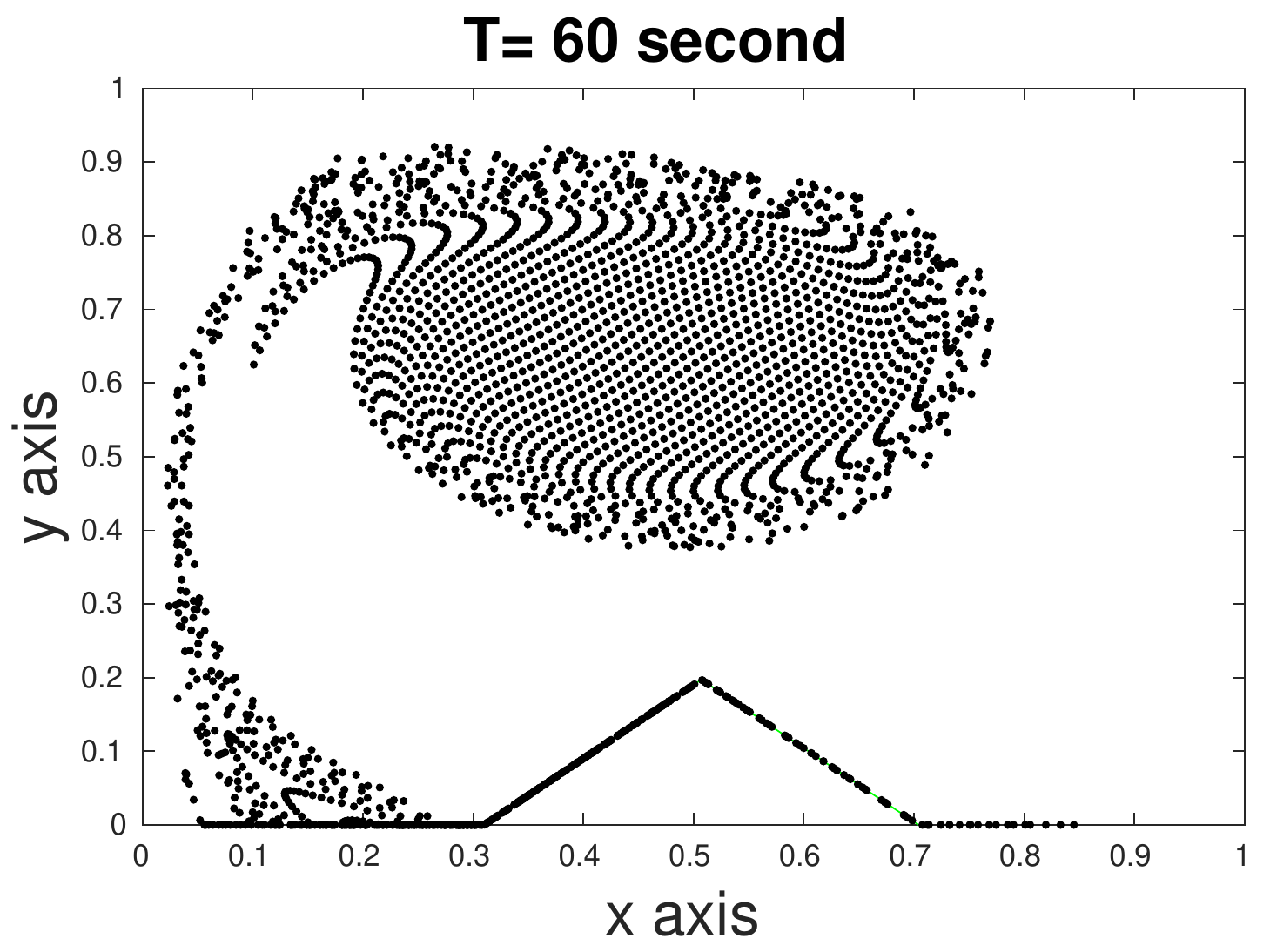}

\centerline{(e)}
\end{minipage}
\hfill
\begin{minipage}{.48\textwidth}
  \includegraphics[width=\linewidth]{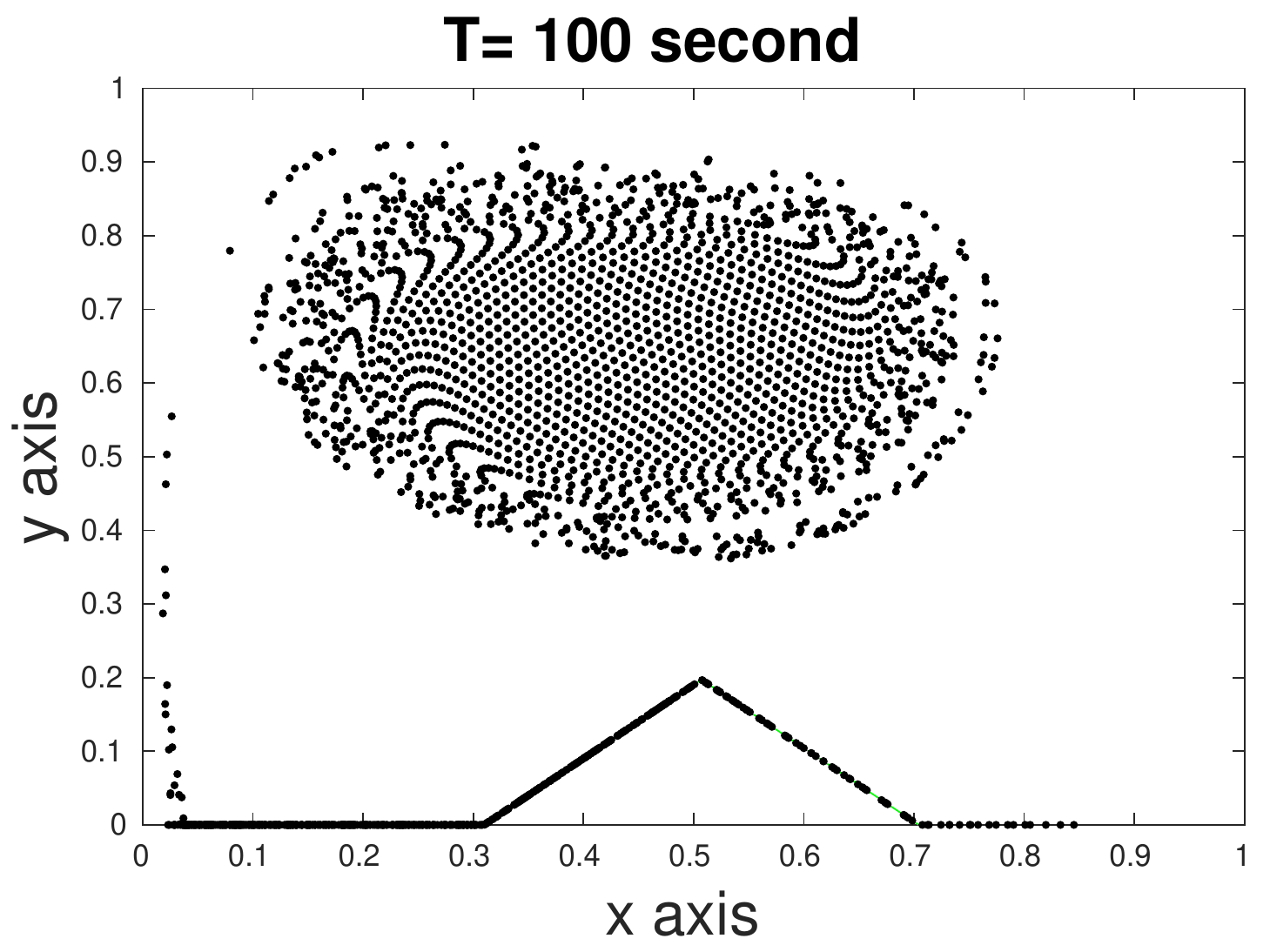}

\centerline{(f)}
\end{minipage}

\caption{Particles position at different times without attachment force on the TGO.}
\label{fig:6}
\end{figure}

In Figure \ref{fig:6}, the sedimentation of the particles on the TGO is shown at different times ignoring the attachment force. The initial size and position are considered same as the above cases. Here the similar physical phenomena are observed, i.e., the rectangular arrangement of particles changes into the circular arrangement with time. From Table \ref{tab:4}, it can be seen that the number of particles sediment on the TGO is more than the OAO. In this case, the particles sedimentation is observed on both the edge and the left side of the TGO. It is noticed that few particles sedimented on the right side of the obstacle, the same as observed in the case of AAO.

\begin{table}[h!]
\caption{Initial condition and parameters}
\label{tab:2}
\begin{adjustbox}{width=1\textwidth}
\begin{tabular}{ p{3 cm} p{3 cm} p{3 cm} p{3 cm}}
 \hline
   & $\rho_{p}$ &$r_{p}$ & Initial position \\
 \hline
\multirow{2}{*}{AAO} & 1030 [$\mathrm{kg}\,\mathrm{m}^{-3}$] & & \\
& 1050 [$\mathrm{kg}\,\mathrm{m}^{-3}$] & $1\times 10^{-4}$ [m] & $(0.3,0.95)$ \\
& 1070 [$\mathrm{kg}\,\mathrm{m}^{-3}$] & & \\
 \cline{2-4}
\multirow{2}{*}{OAO} & 1030 [$\mathrm{kg}\,\mathrm{m}^{-3}$] & & \\
& 1050 [$\mathrm{kg}\,\mathrm{m}^{-3}$] & $1\times 10^{-4}$ [m] & $(0.3,0.95)$\\
& 1070 [$\mathrm{kg}\,\mathrm{m}^{-3}$] & & \\
 \hline
\end{tabular}
\end{adjustbox}
\end{table}

\begin{figure}[h!]
\begin{minipage}{.48\textwidth}
  \includegraphics[width=\linewidth]{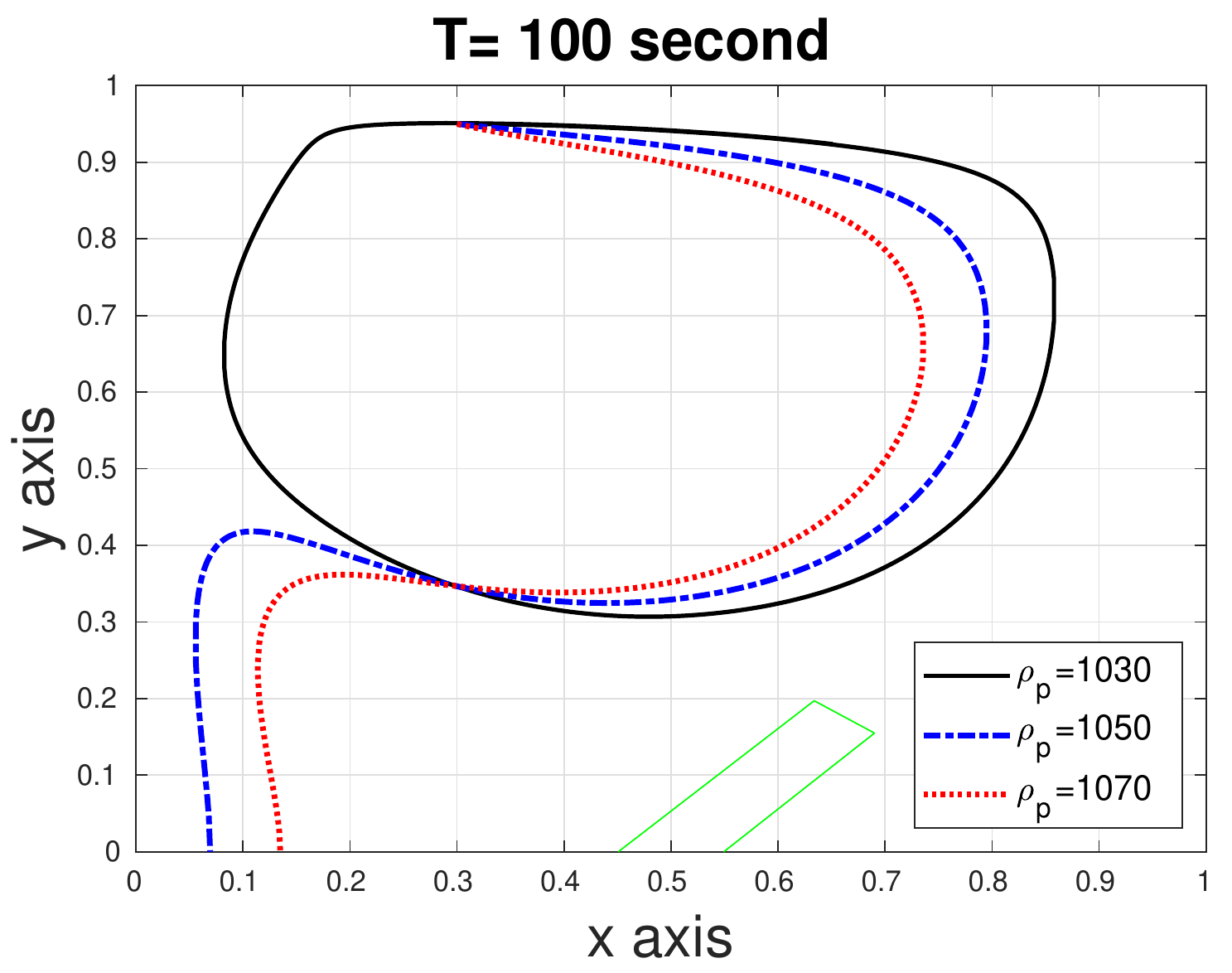}
\caption{Particle sedimentation on the AAO for different densities.}
\label{fig:7}
\end{minipage}
\hfill
\begin{minipage}{.48\textwidth}
  \includegraphics[width=\linewidth]{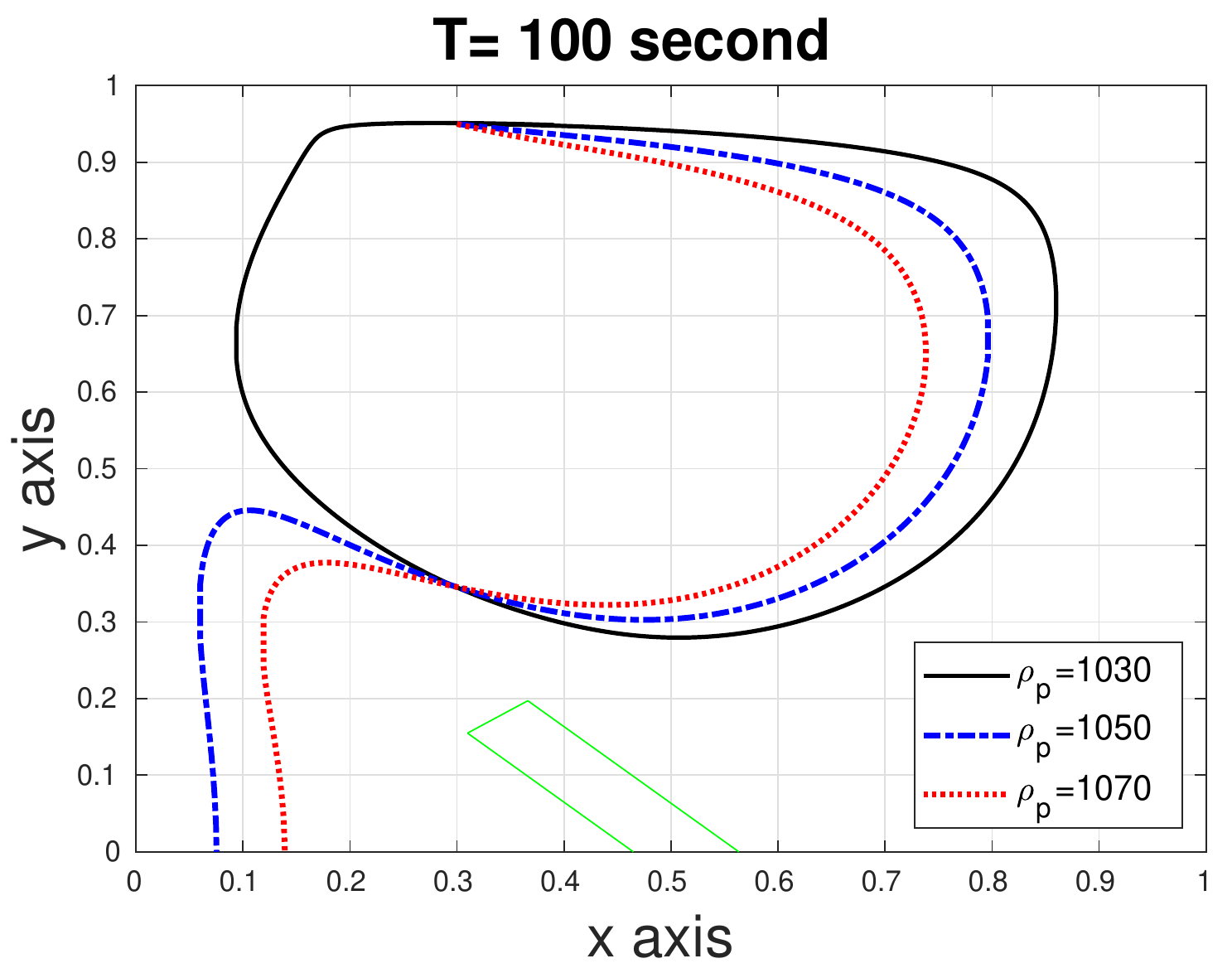}
 \caption{Particle sedimentation on the OAO for different densities.}
\label{fig:8}
\end{minipage}
\end{figure}

\par To show the impact of particle densities on the sedimentation of particle, we considered single particle, see Figures \ref{fig:7} and \ref{fig:8}.  The initial condition and other parameters are shown in Table \ref{tab:2}. It can be seen that the high-density particle had faster sedimentation rate compared to low-density particles. Furthermore, the same phenomena can be depicted for different radius particles. In particular, particle mass can be increased to increase the density of the particle. From here, one can conclude that the trajectory of the light-weight particle follows the fluid primary vortex spiral. But the heavy particle does not follow the fluid streamline because the resultant drag force is not enough to drift the particle, which can be seen from Figures \ref{fig:7} and \ref{fig:8} , respectively. Also, the particle trajectories are prevailed by the gravitational force, where the fluid velocity is very less. The different settling positions are observed with the different initial positions of the particle. The sedimentation highly depends on the initial position of the particle. The less percentage of the particles are settling when the initial position is closer to the center of the primary vortex. The particle is moving in a circular trajectory if it is placed nearer to the center of the primary vortex.

\subsection{Influence of the attachment force}
\begin{figure}[h!]
\begin{minipage}{.48\textwidth}
  \includegraphics[width=\linewidth]{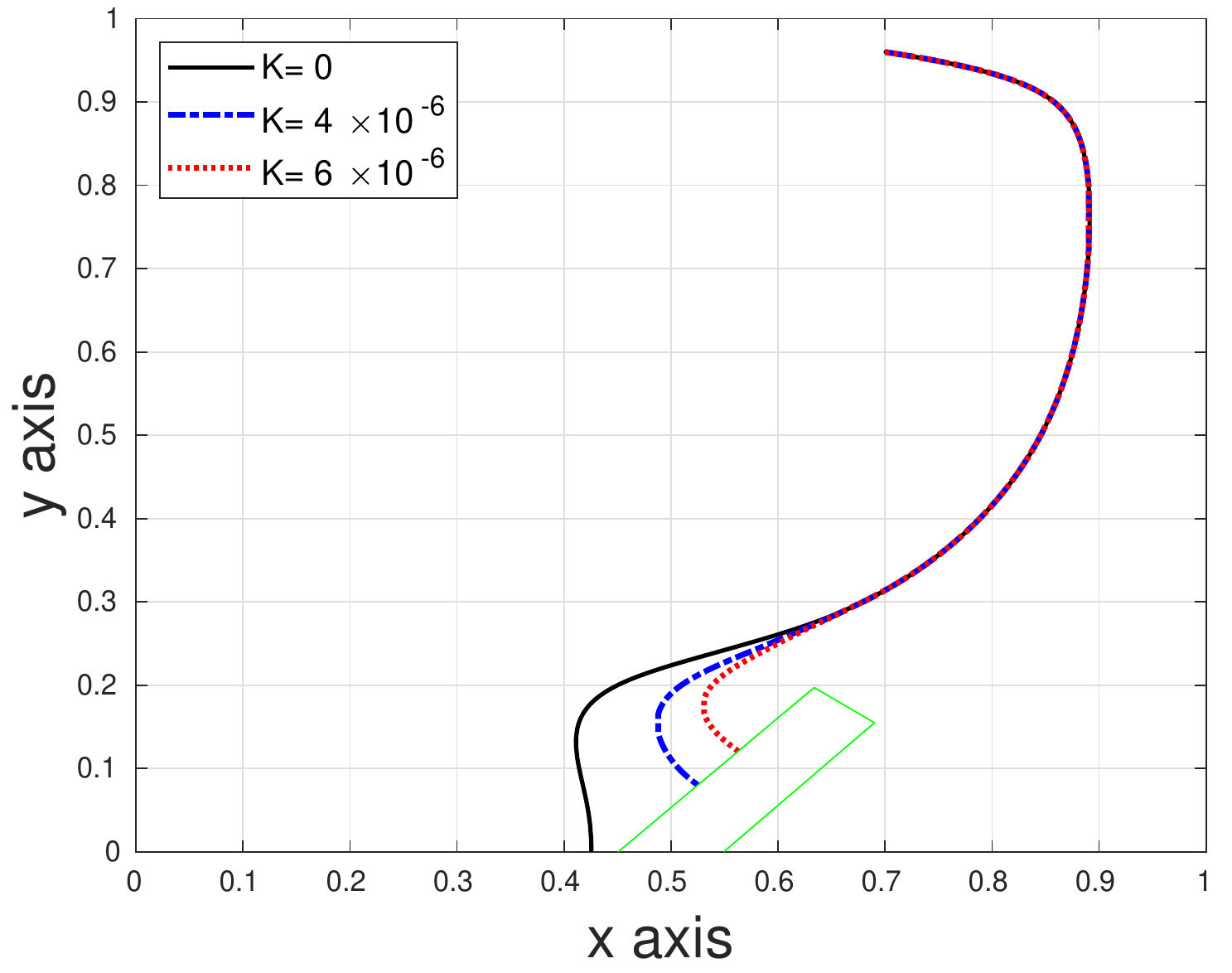}
\caption{Particle sedimentation with and without attachment force on the AAO.}
\label{fig:9}
\end{minipage}
\hfill
\begin{minipage}{.48\textwidth}
  \includegraphics[width=\linewidth]{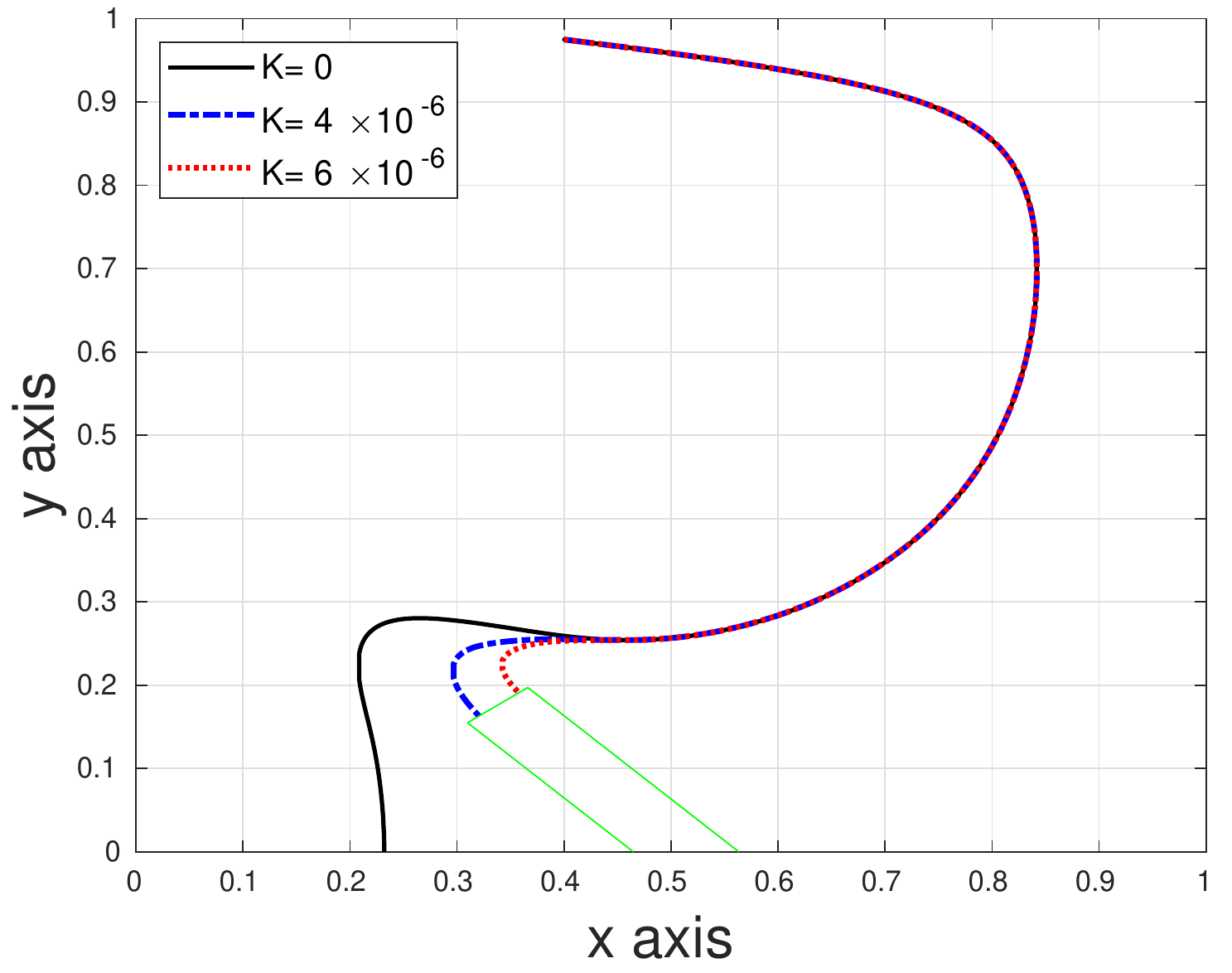}
\caption{Particle sedimentation with and without attachment force on the OAO.}
\label{fig:10}
\end{minipage}
\end{figure}

 It is seen that the sedimentation of the particle depends on various things, such as, particles' initial position, size, density, the velocity of the fluid and the geometry of the cavity, etc. To exhibit the impact of the attachment force on the sedimentation of the particles, single particle sedimentation is considered with and without attachment force on the obstacles. We considered three different values of the parameter $K$ to analyze the impact of the attachment force on the particle sedimentation. Here, $K=0$ signifies that there is no attachment force and similarly non zero values of $K$ means there is attachment force involved. The particle size and the other parameters are represented in Table \ref{tab:3}, and also the parameter $\delta$ is considered as shown in Table \ref{tab:1}. The trajectories of the particle with and without attachment force on the AAO is shown in Figure \ref{fig:9}. The Figure \ref{fig:10} shows particle trajectories in case of OAO. It can be seen that once a particle comes inside the region of influence, it is immediately attracted by the obstacle. It is clear that due to the influence of the attachment force, the particle comes near the obstacle. From the above, it can be concluded that the attachment force is responsible for the particle sedimentation on the obstacle. The impact of the attachment force can be increased by increasing the value of $K$ as shown in the Figures \ref{fig:9} and \ref{fig:10}.  
 
\begin{table}[h!]
\caption{Initial condition and parameters}
\label{tab:3}
\begin{adjustbox}{width=1\textwidth}
\begin{tabular}{ p{3 cm} p{3 cm} p{3 cm} p{3 cm} p{3 cm}}
 \hline
   & $K$ &$r_{p}$ &Initial position &$\rho_{p}$\\
 \hline
\multirow{2}{*}{AAO} & $4\times 10^{-6}$ & $1\times 10^{-4}$ [m]&$(0.7,0.96)$ &1050 [$\mathrm{kg}\,\mathrm{m}^{-3}$]\\
& $6\times 10^{-6}$ & & &\\
 \cline{2-5}
\multirow{2}{*}{OAO} & $4\times 10^{-6}$ & $1\times 10^{-4}$ [m]&$(0.4,0.975)$ &1050 [$\mathrm{kg}\,\mathrm{m}^{-3}$]\\

& $6\times 10^{-6}$ & & &\\
 \hline
\end{tabular}
\end{adjustbox}
\end{table}

\begin{figure}[h!]
\begin{minipage}{.48\textwidth}
  \includegraphics[width=\linewidth]{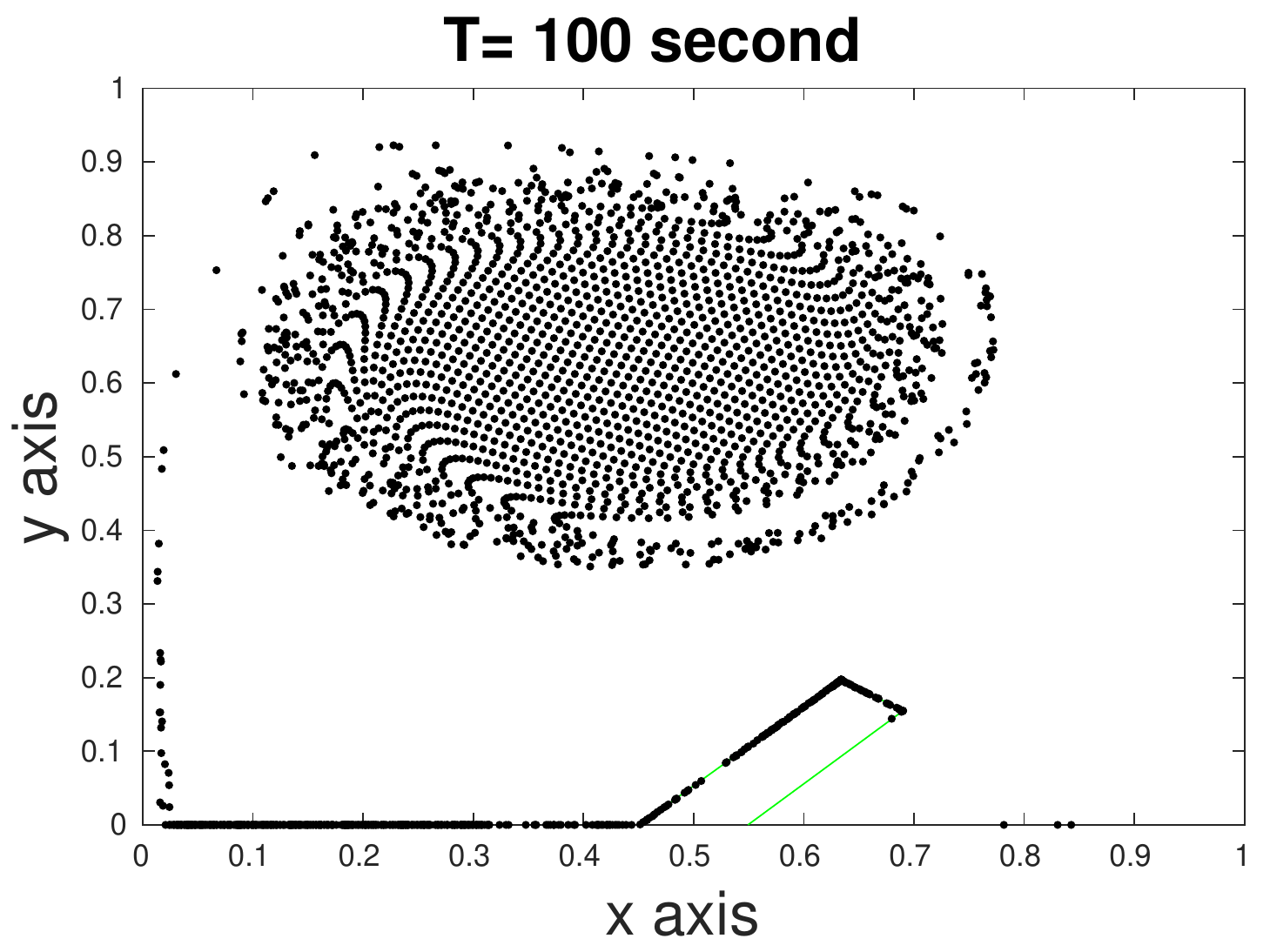}
\end{minipage}
\hfill
\begin{minipage}{.48\textwidth}
  \includegraphics[width=\linewidth]{rotate16.pdf}
\end{minipage}
\caption{Sedimentation of particles with (left) and without (right) attachment force on the AAO.}
\label{fig:11}
\end{figure}

\begin{figure}[h!]
\begin{minipage}{.48\textwidth}
  \includegraphics[width=\linewidth]{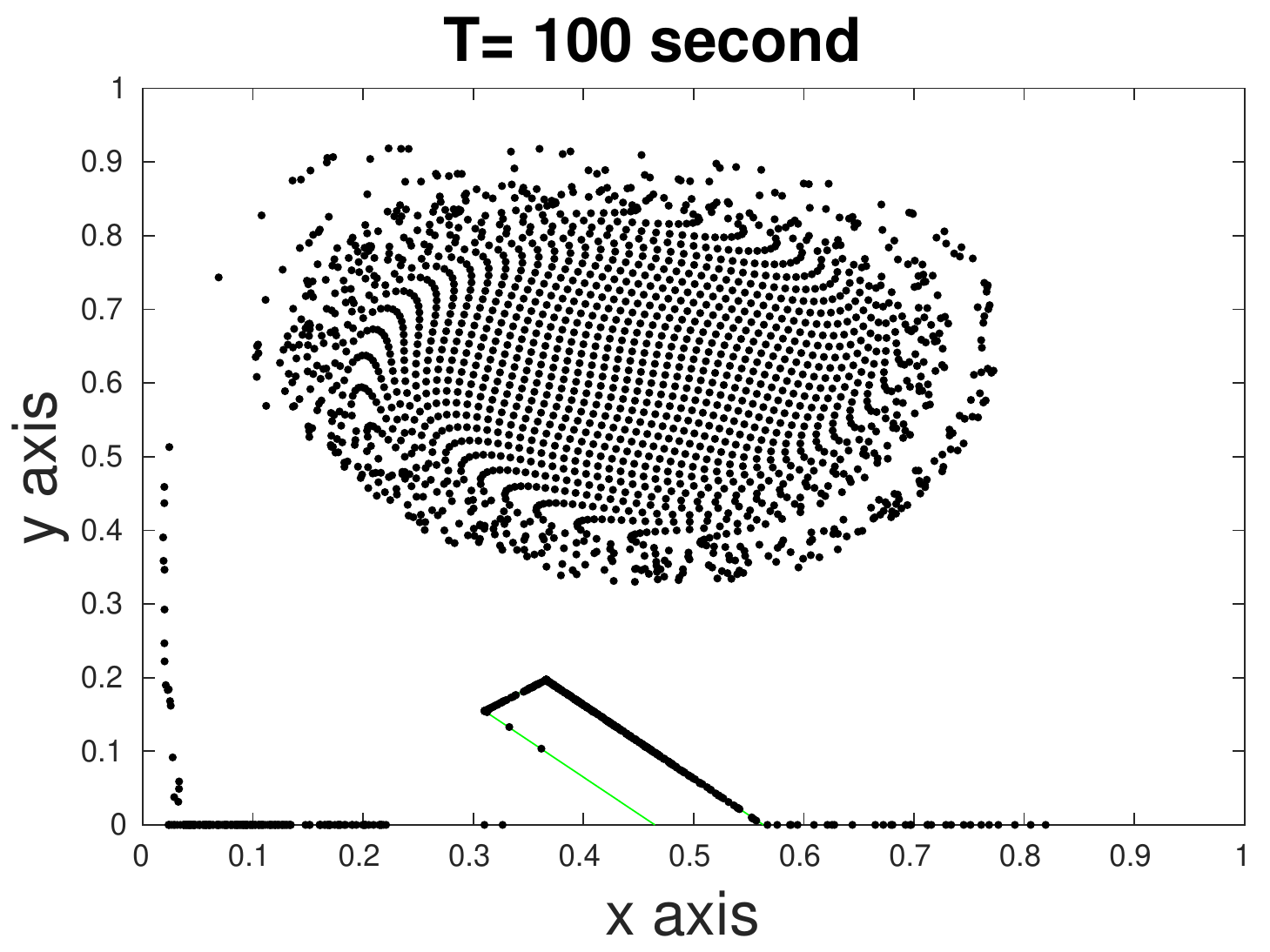}
\end{minipage}
\hfill
\begin{minipage}{.48\textwidth}
  \includegraphics[width=\linewidth]{angle16.pdf}
\end{minipage}
\caption{Sedimentation of particles with (left) and without (right) attachment force on the OAO.}
\label{fig:12}
\end{figure}

\begin{figure}[h!]
\begin{minipage}{.48\textwidth}
  \includegraphics[width=\linewidth]{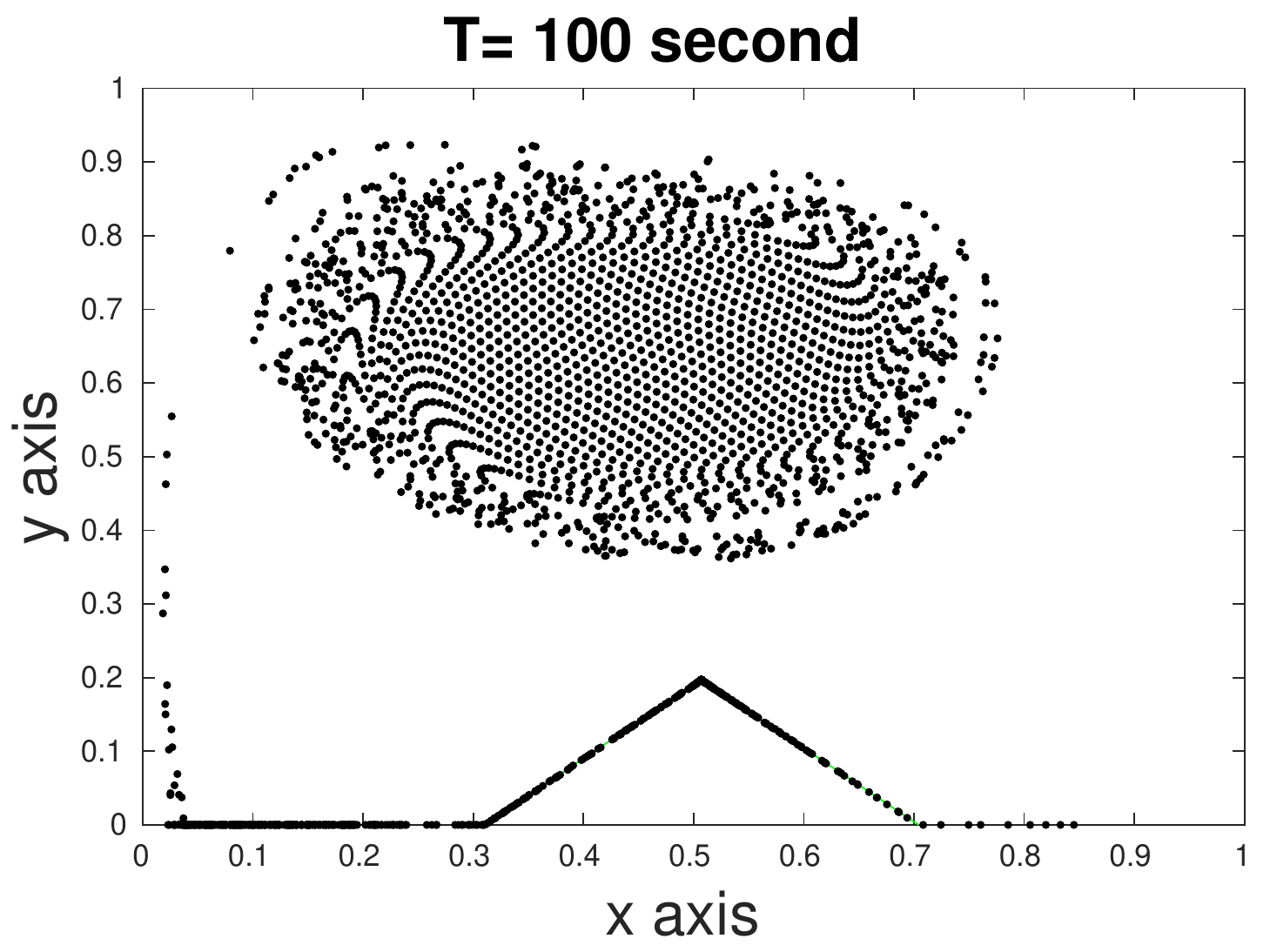}
\end{minipage}
\hfill
\begin{minipage}{.48\textwidth}
  \includegraphics[width=.9\linewidth]{two66.pdf}
\end{minipage}
\caption{Sedimentation of particles with (left) and without (right) attachment force on the TGO.}
\label{fig:13}
\end{figure}

\par The number of particles on the obstacle might be different for different attachment force. Hence, it is requisite to count the number of the particles on the obstacles for different values of $K$. The sedimentation of 2500 particles of size $1\times10^{-4}$[m] with and without attachment force is shown in Figure \ref{fig:11} on the obstacle AAO. All the necessary parameters are considered according to Table \ref{tab:1}. Initially all the particles placed uniformly in the rectangular box $[0.2, 0.9]\times[0.4, 0.9]$. In this case all the results are carried out for time $T= 100s$. Having a closer look at the Figure \ref{fig:11}, we can see that the number of particles on the obstacle \emph{with} the attachment force is much more than  without it. This is due to the attachment force between particles and obstacle wall. The same phenomena are observed for obstacles OAO and TGO, which are shown in Figures \ref{fig:12} and \ref{fig:13}, respectively.

\subsection{Number of sedimented particles}

\begin{figure}[h!]
  \centering
  \includegraphics[width=.7\linewidth]{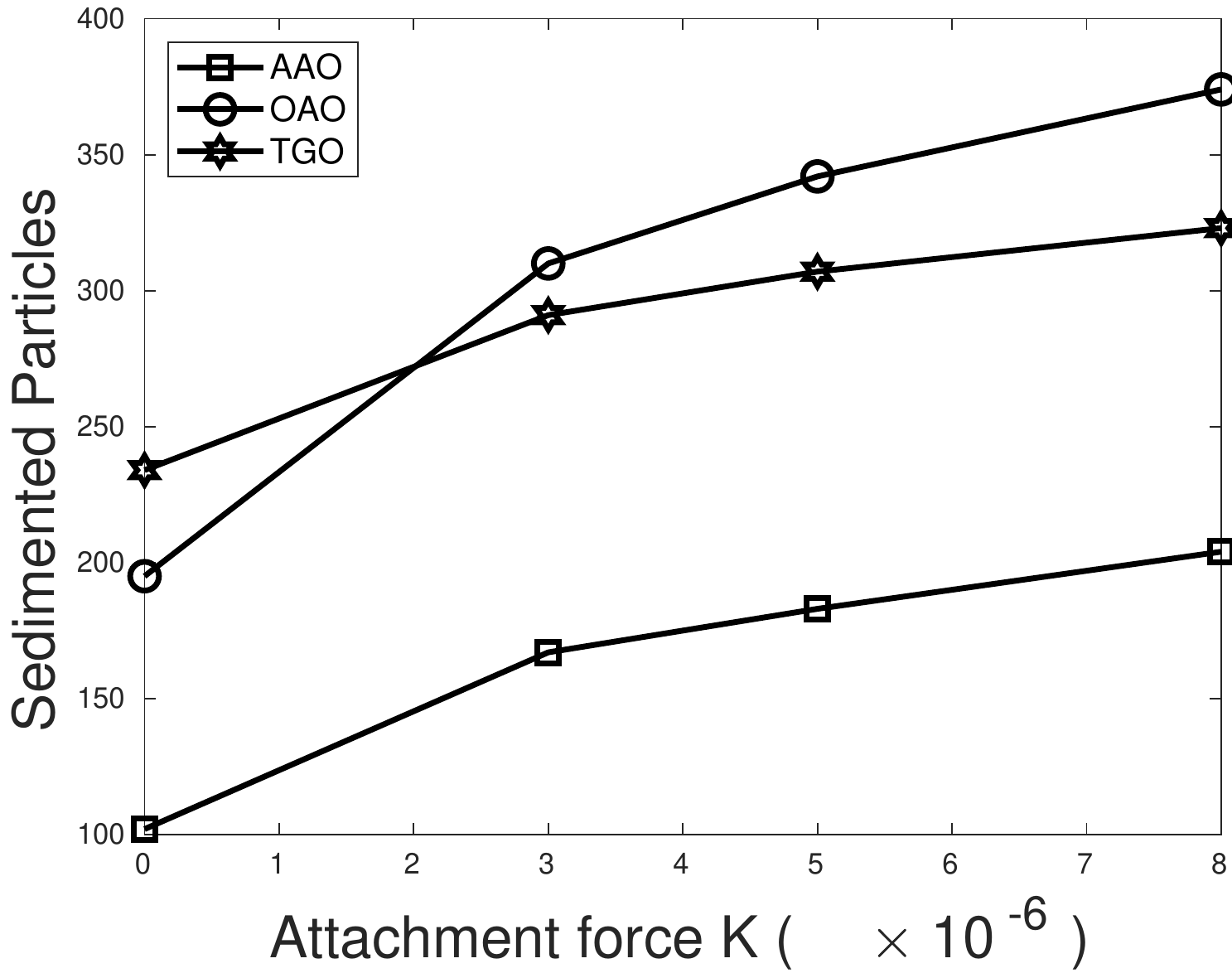}
\caption{Graphical representation of Table \ref{tab:4}.}
\label{fig:14}
\end{figure}

From Table \ref{tab:4}, it can be seen that the number of particles sedimented in case of TGO with no attachment force is more than that of in AAO and OAO. 
The number of sedimented particles increases by increasing the attachment coefficient $K$, as shown in Figure \ref{fig:14}.

\begin{table}[h!]
\caption{Number of sedimented particles on different obstacles with and without attachment force at time $T=100 s$.}
\label{tab:4}
\begin{adjustbox}{width=1\textwidth}
\begin{tabular}{ p{3cm} p{2.5 cm} p{2.5 cm} p{2.5 cm} p{2.5 cm}  }

 \hline
 $K \rightarrow $ &0 & $3\times10^{-6}$ &$ 5\times 10^{-6}$ &$8\times 10^{-6}$\\
 \hline
AAO & 102 & 167 & 183 & 204\\
\hline
OAO & 195 & 310 & 342 & 374\\
\hline
TGO & 234 & 291 & 307 & 323\\
 \hline
\end{tabular}
\end{adjustbox}
\end{table}

\section{Conclusions}
Numerical simulation of the spherical solid particles' behavior in the lid-driven cavity with the presence of different tilted obstacle has been discussed here. This study determines how attachment force and particle density impact the sedimentation behavior of the particles on the obstacle. The significant effect of the buoyancy, gravity, drag, and attachment force in the particles' translational motion have been summarized. It is seen that particles come near the obstacle due to the influence of the attachment force. The sedimentation also depends proportionally on the constant value $K$, i.e., the sedimentation rate on the obstacle will increase by increasing the value of $K$. Here, we have only considered spherical particles. However, the particles in real scenarios may have some other geometry like ellipsoidal, fibers, etc. It might be interesting to consider different shapes of the particle in future studies.

\bibliographystyle{unsrt}
\bibliography{particles_rgsw_arxiv.bib}

\end{document}